\documentclass{jpp}
\usepackage{xcolor}
\usepackage{amsmath}
\usepackage{amssymb}
\usepackage[multidot]{grffile}
\usepackage{graphicx}
\usepackage{epstopdf, epsfig}
\usepackage{hyperref}
\usepackage[utf8]{inputenc}
\newcommand{\iotab}{\lower3pt\hbox{$\mathchar'26$}\mkern-7mu\iota}

\graphicspath{{./}
	{./figsPng/}
}
\shorttitle{Nonlinear PIC simulations in stellarators}
\shortauthor{E. S\'anchez et al.}
\title{Nonlinear gyrokinetic PIC simulations in stellarators with the code EUTERPE}
\author{E. Sánchez\aff{1} 
  \corresp{\email{edi.sanchez@ciemat.es}},
  A. Mishchenko\aff{2},   J.M. García-Regaña\aff{1}, R. Kleiber\aff{2}, A. Bottino\aff{3}, L. Villard\aff{4} and the W7-X team}% and A. Alonso\aff{1}}

\affiliation{
	\aff{1}Laboratorio Nacional de Fusión-CIEMAT, Avda. Complutense 40, 28040, Madrid, Spain.
	\aff{2}Max-Planck Insitut für Plasmaphysik, D-17491 Greifswald, Germany.
	\aff{3}Max-Planck Insitut für Plasmaphysik, D-85748 Garching, Germany.
	\aff{4}Ecole Polytechnique Fédérale de Lausanne, Swiss Plasma Center, CH-1015 Lausanne, Switzerland.
}

\begin{document}
%***************************************************************************************

\maketitle

%***************************************************************************************
\begin{abstract}
	In this work, the first nonlinear particle-in-cell  simulations  carried out in a stellarator with the global gyrokinetic code EUTERPE  using realistic plasma parameters are reported. Several studies are conducted with the aim of enabling reliable  nonlinear simulations in stellarators with this code. First, EUTERPE  is benchmarked against ORB5  in both linear and nonlinear settings in a tokamak configuration. Next, the use of noise control and stabilization tools, a Krook-type collision operator, markers weight smoothing and  heating sources  is investigated. It is studied in detail how these tools influence the linear growth rate of instabilities in both tokamak and stellarator geometries and their influence on the linear zonal flow evolution in a stellarator. Then, it is studied how these tools allow improving the quality of the results in a set of nonlinear simulations of electrostatic turbulence in a stellarator configuration.  
	Finally, these tools are applied to a W7-X magnetic configuration using experimental plasma parameters.
\end{abstract}
%
%
%***************************************************************************************
\section{Introduction}
%***************************************************************************************
Particle-in-cell (PIC) simulations of plasma turbulence have the important drawback that the numerical noise increases with the simulation time. This numerical noise is not critical for linear simulations because the 
linearly unstable modes usually grow much faster than the noise. However, the quantification of physically relevant quantities in a turbulent state requires nonlinear simulations covering a simulation time beyond the nonlinear saturation of linearly unstable modes.
Statistically significant averages of physical quantities can only be extracted in  steady state. Noise accumulation has several deleterious effects in nonlinear simulations. Not only it does reduce the accuracy of any measured quantity but it also makes difficult to reach quasi-steady conditions. A side effect of the numerical noise is the generation of a spurious unphysical contribution to the zonal flow (ZF) component of the turbulent potential. 

This numerical noise in particle-in-cell simulations is known to grow due to the increase in the variance of the markers weights \citep{Krommes1999}.
 As the variance of the weights decreases with the number of markers used in the simulation, a brute force solution consists of reducing the noise level by increasing the number of markers, which, however, increases notably the computational cost and, furthermore, it does not provide a steady state, but only a slowly decaying one. Alternative solutions have been proposed for mitigating this problem, such as the use of a Krook-type operator which introduces a long-time decay of the markers weights \citep{Krommes1999} thus compensating their secular growth, or the use of coarse-graining techniques \citep{Brunner1999} and smoothing of the markers weights \citep{sonnendrucker2015}. Both smoothing and the Krook operator approaches were successfully implemented in the tokamak code ORB5 \citep{McMillan2008}. The Krook operator was shown to have an important effect on the large-scale zonal flow components.  An {\it{ad hoc}} corrected Krook operator was also implemented in ORB5 \citep{McMillan2008}, which preserves the zonal flow residual close to the Rosenbluth-Hinton level in tokamak configurations. However, in stellarators, the linear evolution of zonal flows is more complicated than in tokamaks \citep{Mishchenko2008,Monreal2016,Monreal2017}. The residual zonal flow level is, in general, zero in stellarators and a low frequency characteristic oscillation appears, which makes the implementation of such a correction more difficult than in tokamaks. In addition, the use of a Krook operator in collisional simulations can be questionable, as it can distort the effect of a more realistic collision operator.

 In this work, we report on the effort carried out towards reliable nonlinear simulations with EUTERPE, a global PIC code specifically designed for stellarators \citep{Jost2001}.  We first compare linear and nonlinear simulations with ORB5 and EUTERPE using ideal profiles and a tokamak equilibrium. Then, we study how the weight smoothing, heating sources and the simple Krook operator implemented in EUTERPE, to which we refer generically as noise control and stabilization tools (NCSTs), can improve the quality of the simulations and their effect on the zonal flow evolution in a set of linear and nonlinear simulations in stellarator configurations. These tools are shown to allow reducing the numerical noise at the expense of affecting the zonal flow. Finally, we present the application of these tools to a nonlinear gyrokinetic simulation in a W7-X configuration using realistic experimental plasma parameters.

The rest of the paper is organized as follows. In Section \ref{secCodes} the codes EUTERPE and ORB5 and the equations they solve are presented. In Section \ref{secEUT-ORB5_Bench} a comparison of simulations carried out with both codes is presented. In Section \ref{secNCSTsCharStell} a detailed characterization of noise control and stabilization tools is presented. Section \ref{secApplW7XReal} is devoted to present an application to a realistic plasma from W7-X in which these tools are used to stabilize and improve the quality of the simulation. Finally, in Section \ref{secSumConcl} a summary is presented and some conclusions are drawn.

%***************************************************************************************
\section{The codes EUTERPE and ORB5}\label{secCodes}
%***************************************************************************************
     
	EUTERPE \citep{Jost2000a} and ORB5 \citep{Tran99} share a common origin. They were both initially developed at CRPP Lausanne. Both are particle-in-cell $\mathrm{delta}-f$ codes and many features first developed for one of them were afterwards implemented into the other one, as a result of the close collaboration of the corresponding development groups at Greifswald and CRPP/Garching, respectively.
	However, they also have significant differences, the most important one of which is that ORB5 handles only axysimmetric equilibrium, whereas EUTERPE was designed from the beginning for stellarators and can treat any magneto-hydrodynamic (MHD) equilibrium calculated with VMEC, either tokamak or stellarator.
	Both codes also differ in the equilibrium distribution function that they use. EUTERPE uses a local Maxwellian, while ORB5 can use either a local or a canonical Maxwellian \citep{Angelino2006,vernay_global_2010}. ORB5 is also more flexible in the scheme used for the time evolution of the distribution function, it being able to use a direct $\mathrm{delta}-f$ scheme  \citep{Allfrey2003} or a standard $\mathrm{delta}-f$ one, while in EUTERPE only the standard $\mathrm{delta}-f$ scheme has been implemented so far. More details on the current status of the ORB5 code can be found in \citep{lanti_orb519}.

	As a matter of fact, the applications of ORB5, restricted to the tokamak domain, have been more extensive than those of EUTERPE and many nonlinear turbulence tokamak simulations with ORB5 have been reported (see for instance \citep{Villard19} and references therein). In the case of EUTERPE, nonlinear electrostatic simulations have only been studied in a screw pinch geometry so far \citep{Sanchez2010}. 
	
	Both codes solve the gyrokinetic equations derived from a variational principle \citep{Hahm1988a,Tronco2016}.
	In this work we restrict to the electrostatic case, although both codes have implemented equations for the evolution of the electromagnetic  potential.
	
	An equation for the evolution of the distribution function of each kinetic species, $a$, is solved 
	\begin{equation}
		\frac{\partial f_a}{\partial t}+ \dot{\mathbf{R}} \frac{\partial f_a}{\partial \mathbf{R}}  + \dot{v_{\|}} \frac{\partial f_a}{\partial v_{\|}} = \sum_b C(f_a,  f_b) + S,
	\end{equation}
	where $S$ here represents the sources that can be added with the purpose of controlling the numerical noise and/or sustaining density and temperature profiles. $S=S_{ka} + S_{ha}$, with $S_{ka}$ being a Krook operator and $S_{ha}$ a heating source term. 
	A Krook operator, including corrections for energy, momentum and ZF conservation, is implemented in ORB5 \citep{McMillan2008} while in EUTERPE only a simple Krook operator without corrections is implemented.
		
	In this work, collisionless simulations (with $\sum_b C(f_a,  f_b)=0$) are analyzed, except those in Section \ref{secApplW7XReal}, for which a pitch angle scattering collision operator is used \citep{Kauffmann2010,Regana2013}.

	The equations of motion, $\dot{\mathbf{R}}$ and $\dot{ v_{\|} }$,  are:
	\begin{eqnarray*}
		\dot{\mathbf{R}}&=& \underbrace{ v_{\|} \mathbf{b} +  \frac{\mu B  + v_{\|}^{2}} {B^*_{a} \Omega_a} \mathbf{b} \times \nabla \mathbf{B} 
			+ \frac{v_{\|}^2  } {B^*_{a} \Omega_a} (\nabla\times \mathbf{B})_{\perp}  - {  \frac{ \nabla  \phi _{ext}  \times \mathbf{b} }	{B^*_{a}}}
			 }_{\dot{\mathbf{R}}^0} 
		{\underbrace{- \frac{ \nabla \langle \phi \rangle  \times \mathbf{b}  }{B^*_{a}}}_{\dot{\mathbf{R}}^1}},\\
		\dot{ v_{\|} } &=& \underbrace{- {\mu} \left[ \mathbf{b}
			+  \frac{v_{\|}}  {B^*_{a} \Omega_a}(\nabla \times \mathbf{B})_{\perp}
			\right] \cdot \nabla \mathbf{B}}_{\dot{ v_{\|} }^0}
		{ \underbrace{-  \frac{q_{a}} {m_{a}} \left[ \mathbf{b}   +  \frac{ v_{\|} }  {B^*_{a} \Omega_{a}} \left(\mathbf{b} \times \nabla \mathbf{B} +  (\nabla \times \mathbf{B})_{\perp} \right)\right]
				\cdot \nabla \langle \phi \rangle }_{\dot{ v_{\|} }^1}} ,
	\end{eqnarray*}		
	where $\mathbf{R}$ is the position of the gyrocenter and $v_{||}$ the parallel velocity, and the upper dot means time derivative. The magnetic moment, $\mu$, is a constant of motion ($\dot{\mu } = 0$); 
	$q_a$ and $m_a$ are the charge and mass, respectively, of the species $a$;
	$\Omega_a=\frac{q_aB}{m_a}$, and 
	$B^*_{a}= B + \frac {m_a v_{||}}{q_a} \mathbf{b} \cdot \nabla \times \mathbf{b}$. The fourth term in the right hand side of the equation for $\dot{\mathbf{R}}$ represents the contribution of a long-wavelength ambient electric field $\nabla \phi_{ext}$ that can be introduced externally and is kept fix during the simulation, while the fifth one represents that of the perturbed electrostatic potential $\phi$, which is consistently evolved.

	A $\mathrm{delta}-f$ splitting is used in both codes, so that the distribution function is separated into two parts:
	\begin{equation}
	f_a(\mathbf{R}, v_{||}, \mu, t) =  f_{Ma}(\mathbf{R}, v_{||}, \mu)+  \delta f_a(\mathbf{R}, v_{||}, \mu, t),
	\end{equation}
	with  $f_{Ma}$ being a local Maxwellian in EUTERPE, and either a local or a canonical Maxwellian distribution function in  ORB5.

	With this splitting, a general nonlinear equation for the $\mathrm{delta}-f$ can be obtained,
		\begin{equation}
	\frac{\partial \delta f_a}{\partial t} + \dot{\mathbf{R}} \frac{\partial \delta f_a}{\partial \mathbf{R}} +  \dot{ v_{\|}} \frac{\partial \delta f_a}{\partial v_{\|}} = 
	- { \dot{ \mathbf{R}}^1} \frac{\partial f_{Ma}}{\partial \mathbf{R}} - { \dot{ v_{\|}}^1} \frac{\partial f_{Ma}}{\partial v_{\|}} + \sum_b C(f_a,  f_b) + S,
	\label{deltaFEqn}
	\end{equation}
 assuming that $f_{Ma}$ is the equilibrium distribution function, or control variate \citep{Aydemir94}. This equation 
	 can be linearized if $\dot{ \mathbf{R}}^1$ and $\dot{ v_{\|}}^1$ are dropped from $\dot{ \mathbf{R}}$ and $\dot{ v_{\|}}$ in the left hand side of Eq. (\ref{deltaFEqn}). 

	The distribution function is discretized by using a Klimontovich representation with markers, or quasiparticles, which follow the particle trajectories and carry a contribution to the distribution function (weight) whose evolution equation can be obtained from  the equation for  $\delta f$:
	\begin{equation}
	\frac{dw_p}{dt}= \frac{\Omega_p}{N_s} \frac{d \delta f_a}{d t}\biggr\rvert_{Z_p} = \frac{\Omega_p}{N_s} \ \biggr[
	- { \dot{ \mathbf{R}}^1} \frac{\partial f_M}{\partial \mathbf{R}} - { \dot{ v_{\|}}^1} \frac{\partial f_M}{\partial v_{\|}} + \sum_b C(f_a,  f_b) + S \biggr]_{Z_p},
	\end{equation}
	where $\Omega_p$ is the phase space volume associated to the p-th marker, $w_p$ is its weight, $N_s$ is the total number of markers in the simulation and $Z_p$ represents the location of the marker in phase space.

	The system of equations is closed with the quasi-neutrality equation, which for the electrostatic case with adiabatic electrons, and using a long wavelength approximation, reads
	\begin{equation}
		q_i \langle n_i \rangle  - \frac{e n_0 (\phi - \{\phi\}_s)}{T_e}=    - \nabla  \frac{m_in_0}{B^2} \nabla_{\perp} \phi
	\end{equation}
	
	Here, the $\langle  \rangle$ represents a gyro-average and $\{  \}_s$ represents a flux-surface average. 
	
	For more details about the equations solved and their implementation in each code, the reader is referred to \citep{Jost2001,Kornilov2004,Slaby18,Jolliet2007,lanti_orb519}.
%			\item  Quasineutrality (LWA)
%			\vspace{-2.5\baselineskip}
%			\hspace{0cm}
%			\begin{eqnarray*} %\quad \quad  \quad \quad %\quad  \quad
%				%\hspace{-40pt}\rm{ad.} \hspace{5pt} e^{-}& \rightarrow & q_i\langle n_i \rangle - \frac{e n_0 (\phi - \bar{\phi})}{T_e}=    - \nabla  \frac{m_in_0}{B^2} \nabla_{\perp} \phi \\
%				%{\color{blue}\rm{kin.\hspace{2pt} ions ,\hspace{2pt} }e^-} &\rightarrow &{\color{blue}{q_i}\langle n_i \rangle - n_e =  -\nabla  \frac{{m_i}n_0}{B^2} \nabla_{\perp} \phi }\\\quad \quad \quad \quad 
%				%\vspace{5pt}\hspace{-5pt}
%				%{\color{red}\rm{kin.\hspace{2pt} ions ,\hspace{2pt} }e^-+\text{Pad\'e}}
%				\color{red}
%				&  & {q_i}\langle n_i \rangle - \frac{en_0(\phi-\bar{\phi})}{T_e} =   - \nabla  \frac{{m_i}n_0}{B^2} \nabla_{\perp} \phi \quad \quad  \quad% + \nabla \rho_i^2  \nabla_{\perp}\left( q_i \langle n_i \rangle - n_e \right)\\
%				%{\color{red}\rm{+  \hspace{2pt}Pade}\hspace{15pt}}& &
%			\end{eqnarray*}

%***************************************************************************************
\section{Benchmark EUTERPE-ORB5}\label{secEUT-ORB5_Bench}
%***************************************************************************************

The code EUTERPE has been benchmarked in  linear settings against several codes, like TORB \citep{Sanchez2010},
% GYGLES \citep{Fivaz1998}, 
GENE \citep{Helander2015,gorler_intercode_2016}, GYSELA, GKW and ORB5 in \citep{gorler_intercode_2016} and XGC \citep{cole_comp2019}. % and GTC \citep{Wang20}. 
In a turbulence nonlinear setting it has only been compared with TORB for electrostatic simulations in a screw pinch geometry so far \citep{Sanchez2010}.

As a step previous to the nonlinear simulations in stellarator geometry with EUTERPE, we carry out a benchmark against ORB5 in tokamak geometry. We choose a tokamak equilibrium, matching the well-known Cyclone Base Case (CBC) tokamak equilibrium \citep{Dimits2000} at middle radius. The main parameters of this equilibrium are major radius $R=1.7 ~\rm{m}$, minor radius $a=0.625~\rm{m}$, and the rotational transform $q(0)=0.85$ and $q(a)=3.24$.% and the radial profile of rotational transform is shown in figure \ref{fig:qNTProfileskTxknY}. 
%%*************************************************************************************
%\begin{figure}
%	\centering
%{\includegraphics[draft=false, trim=0 0 0 0, clip, width=5cm]{CBC_q_profile.png}}
%	\caption{Radial profile of the rotational transform (q) for the Cyclone Base Case used in the ORB5-EUTERPE benchmark.}
%	\label{fig:qNTProfileskTxknY}
%\end{figure}
%%*************************************************************************************

\begin{figure}
	\centering
	{\includegraphics[trim=10 0 0 0, clip, width=6.5cm]{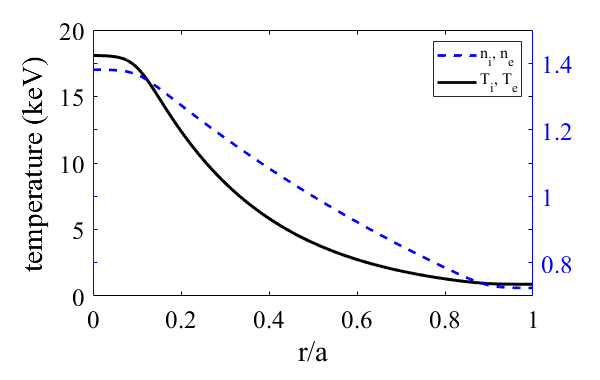}}
	{\includegraphics[trim=10 0 0 0, clip, width=6.5cm]{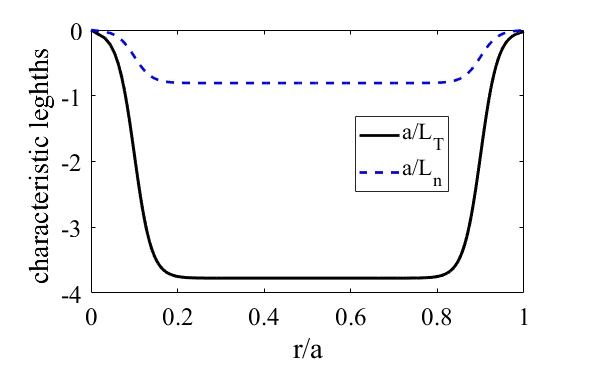}}\\
	\caption{Density and temperature profiles (left) and their characteristics scale-lengths (right) for the analytical profiles defined in Eq. (\ref{analProfilesnT}) with parameters  $\kappa_{Ti} = \kappa_{Te}=3.774$, $\kappa_{ni}=\kappa_{ne}=0.807$, $\Delta_n=\Delta_{Te}=\Delta_{Ti}=0.4$, $\Delta_s=0.04$.}
	\label{fig:profsNLBenchORB5EUT150M}
\end{figure}

A set of linear and nonlinear simulations were carried out with both codes using this MHD equilibrium, adiabatic electrons and ideal density and temperature profiles constructed according to the analytic formula
\begin{eqnarray}
	X=X_* \rm{exp} \left\{ \frac{\kappa_X \Delta_X}{2}  ln \left[ \frac{cosh(\frac{r/a-(\rho_0+\Delta_X)}{\Delta_s})}{cosh(\frac{r/a-(\rho_0-\Delta_X)}{\Delta_s})} \right] \right\},
	\label{analProfilesnT}
%	\label{analProfilesn}
%	T_i,T_e=T_0 \rm{exp} \left\{ \frac{\kappa_{Ti,e} \Delta_{Te,i}}{2} ln \left[ \frac{cosh(\frac{r/a-0.9}{\Delta_{Te,i}})}{cosh(\frac{r/a-0.1}{\Delta_{Te,i}})} \right] \right\}.
%	\label{analProfilesT}
\end{eqnarray}
where $X=\{n_i, n_e,T_e,T_i\}$ represents a radial profile of density or temperature for ions or electrons. $X_*$ is the value of $X$ at the reference normalized radius $\rho_0=0.5$, and $\kappa_X$ and $\Delta_X$ are specific parameters for each profile.
An example of this type of profiles % for the particular case with $\kappa_{Ti} = \kappa_{Te}=3.774$, $\kappa_{ni}=\kappa_{ne}=0.807$, and $\Delta_{ni}=\Delta_{ne}=\Delta_{Te}=\Delta_{Ti}=0.4$, $\Delta_s=0.04$,
 is shown in figure \ref{fig:profsNLBenchORB5EUT150M}.

%%*************************************************************************************
%\begin{figure}
%	\centering
%	{\includegraphics[draft=false, trim=0 0 0 0, clip, width=5cm]{CBC_q_profile.png}}
%	\caption{Analytic density (left) and temperature (right) profiles given by Eqs. (\ref{analProfilesn}-\ref{analProfilesT}) with $\kappa_T=3$, $\kappa_n=0.8$.}
%	\label{qNTProfileskTxknY}
%\end{figure}
%%*************************************************************************************
%*************************************************************************************
%*************************************************************************************

%***************************************************************************************
\subsection{Linear simulations}
%***************************************************************************************

With these model density and temperature profiles we carry out a set of linear simulations for several values of the parameter $\kappa_{Ti}=2,2.54,3,3.5$. In all cases, the electron temperature and density gradients are kept the same with $\kappa_{Te}=0.3775$, $\Delta_s=0.04$, $\kappa_{ni}=\kappa_{ne}=0.8$, $\Delta_{ne}=\Delta_{Te}=0.4$, and  $\Delta_{ni}=\Delta_{Ti}=0.2$. Simulations are global in radius and the maximum instability appears around $r/a\sim 0.38$. The same settings are used in ORB5 and EUTERPE codes: a spatial resolution in the radial\footnote{Note that the radial coordinate, s, is normalized toroidal flux in EUTERPE, while it is the normalized poloidal flux in ORB5}, $s$, poloidal, $\theta_*$, and toroidal, $\phi$, directions $n_s \times n_{\theta_*} \times n_{\phi}=64 \times 128\times 64$ and a squared low pass Fourier filter keeping modes $-63 < m <63$, plus a diagonal filter suppressing modes with $|m - n/\iotab|> 5$. Just a toroidal mode  $n  = 16$ is kept  in both codes. 

\begin{figure}
	\centering
	{\includegraphics[draft=false, trim=10 0 25 0, clip, width=6.5cm]{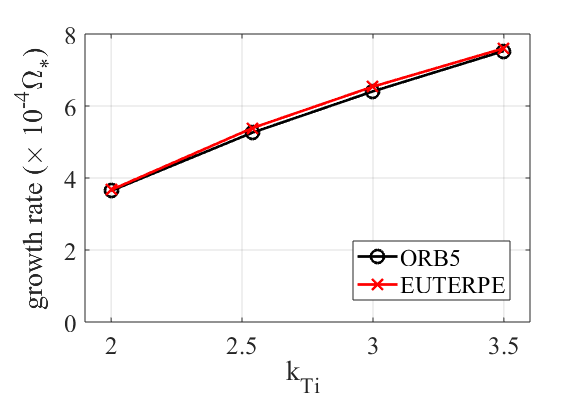}}
	{\includegraphics[draft=false, trim=0 0 25 0, clip, width=6.5cm]{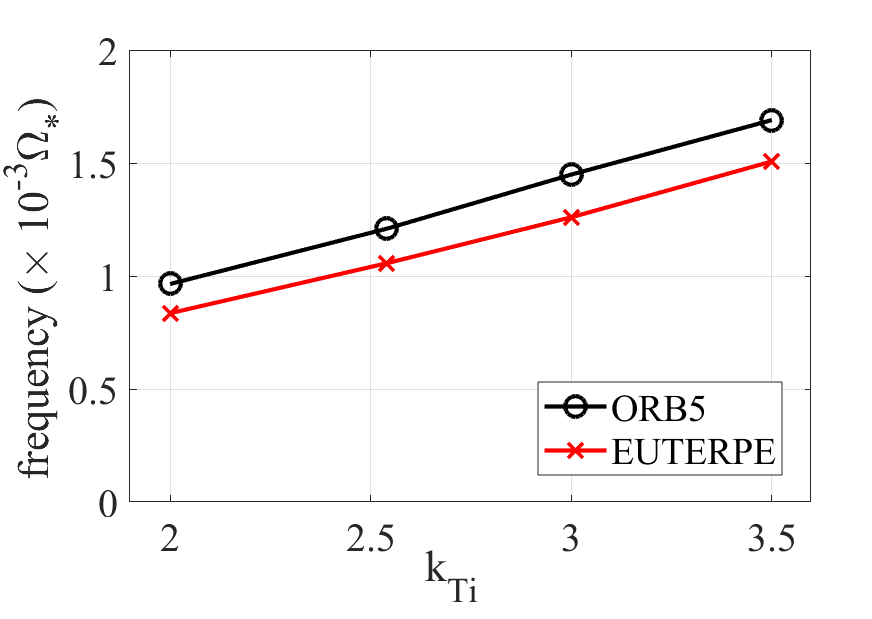}}\\
	\caption{Comparison of growth rate (left) and frequency (right)
		of the most unstable mode in ORB5 and EUTERPE for different values of $\kappa_{Ti}$.}
	\label{ResLinBenchORB5EUT}
\end{figure}

The growth rate and the real frequency are obtained by fitting the time evolution of the potential to an oscillation whose amplitude grows in time. From the fitting, both the growth rate and real frequency of the most unstable mode can be obtained. 
%The results of these fittings for both codes are shown in figure \ref{ResLinBenchORB5EUT}.
%The growth rate is obtained by fitting the time evolution of the potential to an oscillation whose amplitude grows in time. %From the fitting, both the growth rate and real frequency of the most unstable mode can be obtained. 
The results of these fittings for both codes are shown in figure \ref{ResLinBenchORB5EUT}.
%*************************************************************************************
%*************************************************************************************
As shown in the figure, the agreement on the growth rates is excellent, while it is not that good in the real frequencies, which are more sensitive to the precise time evolution of the modes. Note that in these simulations several $m$ modes are present with comparable growth rates and different frequencies. This can affect the fitting to a unique frequency. Also, differences in the code implementation details can affect the frequency results,  while the growth rates appear to be more robust.

%	%
%***************************************************************************************
\subsection{Nonlinear simulations} 
%***************************************************************************************
Once the results of both codes have been successfully compared for linear simulations we proceed with the comparison in a nonlinear setting. 
We use the same CBC magnetic equilibrium and for the density and temperature profiles we use analytic profiles described by Eq. (\ref{analProfilesnT}) with parameters from  \citep{McMillan2008}. The density and temperature profiles are the same for electrons and ions with $\kappa_{Ti} = \kappa_{Te}=3.774$, $\kappa_{ne}=\kappa_{ni}=0.807$ and $\Delta_n=\Delta_{Te}=\Delta_{Ti}=0.04$. These profiles are shown together with their characteristic scale lengths in figure \ref{fig:profsNLBenchORB5EUT150M}. 
	%Simple Krook\\
	%

For this comparison we target the full-volume-integrated heat flux, $Q_i $, as the fundamental physical quantity to study. This quantity is obtained by averaging the radial flux of kinetic energy over all  the markers used in the simulation.
\begin{equation}
	Q_{i}=\sum_{j=1}^{N}(f_{0j}+w_j)\frac{m_j v_j^2}{2} \frac{\left<\vec{E}\right>\times \vec{B}}{BB^*_{\|}}\cdot \frac{\nabla s} {|\nabla s|}
	\label{eqVolIntegHeatFluxDef}
\end{equation}
Here $f_{0j}$ represents the average of the equilibrium distribution function on the $j-\mathrm{th}$ marker's volume and $w_j$ is the contribution to the $\delta f$ (weight) of the $j-\mathrm{th}$ marker.

In addition to the heat flux, the volume-averaged heat conductivity, which is computed as
\begin{equation}
 \chi_i = \frac{\sum_k V_k  \chi_{ik}}{ \sum_k V_k},
 \label{eqHeatDiffsvtyDef}
\end{equation}
  is also compared between both codes. The computation of this magnitude requires a radial binning to obtain the conductivity in a finite number of radial positions, with $V_k$ being the volume of the $k-\mathrm{th}$ radial bin.  

The values of the heat conductivity, $\chi_{ik}$, are obtained by averaging properties of markers within each radial bin, as $\chi_{ik} = Q_{ik} /\nabla T_{ik}$,
with $\nabla T_{ik}$ being the average ion temperature gradient in the $k-\mathrm{th}$ bin and
\begin{equation}		
Q_{ik} = \sum_{s_{k_0} \leq s_j <s_{k_1} }(f_{0j}+w_j)\frac{m_j v_j^2}{2} \frac{\left<\vec{E}\right>\times \vec{B}}{BB^*_{\|}}\cdot \frac{\nabla s} {|\nabla s|},
\label{eqHeatFluxBin}
\end{equation}
where $s_j$ represents the radial location of the $j-\mathrm{th}$ marker and $s_{k_0}$ and $ s_{k_1}$ represent the limits of the $k-\mathrm{th}$ radial bin, centered around the radial position $s_k$.

 In order to get a stable signal for the heat flux we resort to the Krook operator which is now implemented in both codes and has proven to allow the stabilization of noise and heat flux in nonlinear simulations with ORB5 \citep{McMillan2008}. In EUTERPE only the simple Krook operator is implemented, while in ORB5 a corrected-Krook, including energy and momentum conservation corrections,  is also available. Then, for a fair comparison, we switch off these corrections in ORB5 to use the simple Krook version also in it. No additional sources or weight smoothing is used in these simulations.

%*************************************************************************************
\begin{figure}
	\centering
	{\includegraphics[trim=20 0 25 0, clip, width=6.5cm]{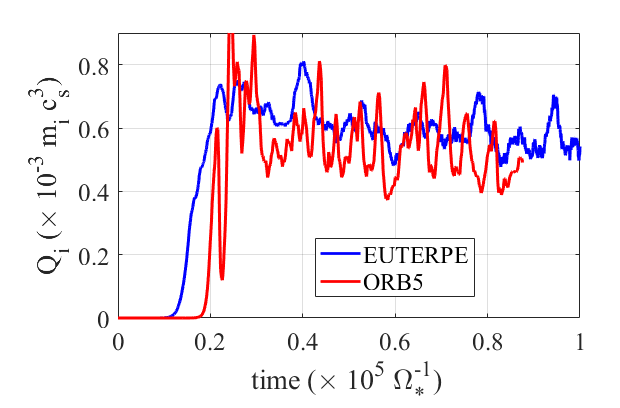}}
	{\includegraphics[trim=20 0 25 0, clip, width=6.5cm]{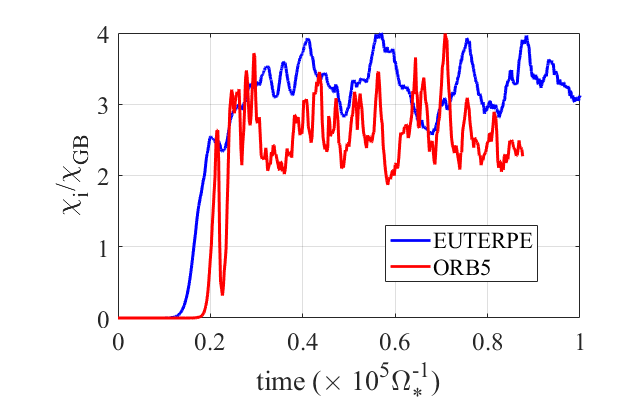}}\\
	\caption{Full-volume integrated heat flux  from Eq. \ref{eqVolIntegHeatFluxDef} (left) and volume-averaged heat conductivity from Eq. \ref{eqHeatDiffsvtyDef} (right) for two simulations carried out with ORB5 and EUTERPE using the same settings (see the text) and using $150\times10^6$ markers. }
	\label{fig:ResNLBenchORB5EUT150M}
\end{figure}
\begin{figure}
	\centering
	{\includegraphics[draft=false, trim=20 0 25 0, clip, width=6.5cm]{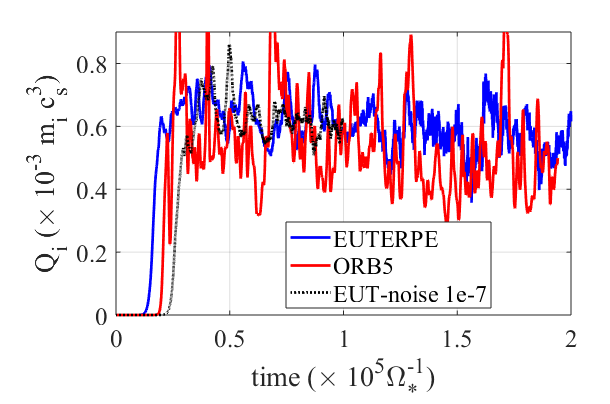}}
	{\includegraphics[draft=false, trim=20 0 25 0, clip, width=6.5cm]{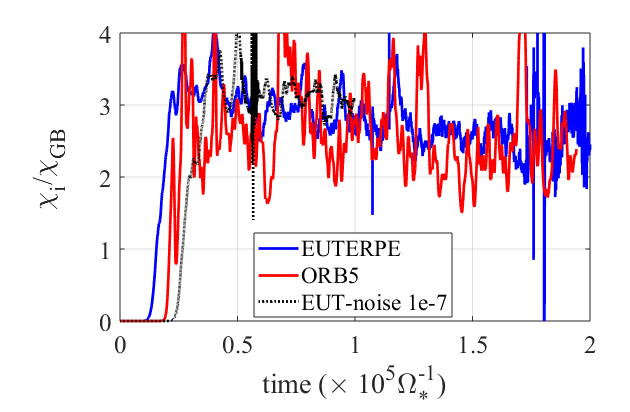}}\\
	\caption{Full-volume integrated heat flux from Eq. \ref{eqVolIntegHeatFluxDef} (left) and volume-averaged heat conductivity from Eq. \ref{eqHeatDiffsvtyDef} (right) obtained in simulations carried out with ORB5 and EUTERPE using the same settings (see the text) and using $450 \times 10^6$ markers. Results of a simulation carried out with EUTERPE using the same initial noise in the markers weights as in ORB5 is also shown in dotted line (labeled EUT-noise 1e-7).}
	\label{fig:ResNLBenchORB5EUT450M}
\end{figure}

%*************************************************************************************
%
	The same input parameters are used in both codes: a spatial resolution $n_s \times n_{\theta_*} \times n_{\phi}=256 \times 256\times 128$, a squared low pass Fourier filter allowing modes $-63 < m <63$, $-63 < n <63$ and a diagonal filter suppressing modes with $|m - n / \iotab  | > 5$; quadratic splines are used in both cases. 
The comparison of the heat fluxes and conductivities is shown in figures \ref{fig:ResNLBenchORB5EUT150M} and \ref{fig:ResNLBenchORB5EUT450M} for two simulations carried out with different number of markers, $150 \times 10^{6}$ and $450 \times 10^{6}$, respectively.  The heat conductivity is normalized to gyro-Bohm units $\chi_{GB}=\rho_i c_s / a$, with $\rho_i$ the ion Larmor radius, $c_s=\sqrt{T_e/m_i}$ the sound speed and $a$ the minor radius.

In the first case, with $150 \times 10^6$ markers, which is shown in figure \ref{fig:ResNLBenchORB5EUT150M}, the statistics is not enough, and the numerical noise is significant, which is manifested in decaying heat flux signals and  translates also into a mismatch between ORB5 and EUTERPE results, particularly in the heat  conductivity, which is affected by both errors in the estimation of heat flux in the radial bins and also by the error in the temperature gradient estimates.

Significantly better agreement between codes is obtained with increased number of markers (figure \ref{fig:ResNLBenchORB5EUT450M}) as compared to the previous one (figure \ref{fig:ResNLBenchORB5EUT150M}). The agreement in the later case is very good in both the heat flux and conductivity, in spite of the implementation details, which suggests that the statistics of $450 \times 10^6$  markers is large enough in this case.

A clear difference between ORB5 and EUTERPE simulations can be observed in the initial stages in figure \ref{fig:ResNLBenchORB5EUT450M}: the exponential growth due to the linear instability initiates first in the EUTERPE case. The difference is due to the fact that a larger amplitude initial noise was used in the EUTERPE case as compared to ORB5 one, and also to the slightly different implementations in both codes.  The results of a simulation carried out with EUTERPE using $450 \times 10^6$ markers and a smaller amplitude initial noise (exactly equal to the value used in ORB5) are also shown in the figure (dotted line). It is clear that the exponential growth of heat flux and conductivity starts later than in the case with larger amplitude noise although an exact match of the growth between both codes is not obtained, due to the different implementations. However, independently of the initial noise used and the different numerical details in both codes, the saturated levels in heat flux and conductivity, which are the physically meaningful quantities, are the same in both codes.

%*************************************************************************************
%*************************************************************************************

%***************************************************************************************
\section{Characterization of noise control and stabilization tools}\label{secNCSTsCharStell}
%***************************************************************************************

In this section, we test the use of the NCSTs %, Krook operator, weight smoothing, and source terms 
and study their influence on the quality and physical results of the simulations.

A simple Krook operator \citep{Krommes1999,McMillan2008} has been implemented in EUTERPE without energy or momentum correction. This simple Krook operator is one of the source terms present in Equation \ref{deltaFEqn} with the form  $S_{ka} = -\gamma_{ka} \delta f_a$, where $\gamma_{ka}$ represents the strength of the source for species $a$.  As in this work we  present simulations with only kinetic ions, in the following sections we will drop the $a$ index related to the kinetic species and $\gamma_k$ will refer to ions. This kind of source term is commonly used to stabilize collisionless simulations. In collisional cases its use is more questionable as it can distort the effect of a more realistic collision operator. In any case, as a common sense rule $\gamma_{k}$ should be chosen small as compared to the growth rate of the unstable modes, and smaller than the collision frequency in collisional simulations, in order to not affect significantly the results. In this section we will show how the size of $\gamma_{k}$ affects the results in different simulations.

In addition to the Krook operator, a weight smoothing scheme \citep{sonnendrucker2015} has also been implemented in EUTERPE. For this purpose, the markers are classified in a quad tree (QT) according to their coordinates in phase space  and the weights of neighbor markers are mixed after a prescribed number of computations, so that their variance is reduced. The weights of a couple of neighboring markers, $w_1$ and $w_2$, are redefined at each smoothing step as:
\begin{eqnarray}
	w_1^{i+1}=(1-e^{-\frac{d^2}{2\sigma}})w_1^i+e^{-\frac{d^2}{2\sigma}}\frac{w_1^i+w_2^i}{2} 
	\label{EqweightChange1}\\
	w_2^{i+1}=(1-e^{-\frac{d^2}{2\sigma}})w_2^i+e^{-\frac{d^2}{2\sigma}}\frac{w_1^i+w_2^i}{2} ,
	\label{EqweightChange2}
\end{eqnarray}
where the $i$ super-index indicates quantities before the redefinition and $i+1$ the quantities after the mixing, and $d^2=(v_{||1}-v_{||2})^2 + (v_{\perp 1}-v_{\perp 2})^2$ represents the separation of markers in velocity space coordinates.

It is clear from the definition of smoothed weights in Eqs. \ref{EqweightChange1}-\ref{EqweightChange2} that after the weights update, the density is conserved and the variance of the weights is reduced. However, with this smoothing scheme neither the energy nor the momentum of markers are conserved. Three parameters can be modified which affect the strength of the weight smoothing. First, the frequency at which the weight smoothing is applied, which we call $f_{QT}$. The second parameter is $\sigma$ in Eqs. (\ref{EqweightChange1} and \ref{EqweightChange2}), which determines the strength of the smoothing and also affects the energy and momentum conservation. The smaller   $\sigma$ is the closer the markers have to be in order to mix their weights. As $\sigma$ increases, the exponential factor in Eqs. (\ref{EqweightChange1} and \ref{EqweightChange2}) increases for large values of the markers separation $d$, and consequently, the conservation of momentum and energy will be worse. In the following we will call this parameter either $\sigma$ or $\sigma_{QT}$. Finally, there is a free parameter in the way the markers are sorted in the quad tree. It is the minimum number of markers which are required in each phase space cell. Note that cells have not the same width but are defined so that each of them contains at least a minimum number of markers per cell, which we call $N_{QT}$. The smoothing is only applied to neighbor markers within the same cell.

These tools have proven to help in mitigating the growth of numerical noise in ORB5, but they can also affect the zonal flow dynamics. While they have shown good results in ORB5 tokamak simulations they have not been  previously tested in a global PIC code in stellarators.

In addition, a heating source term 
\begin{equation}
	S_{ha} = -\gamma_{ta}(\delta f_a(s, \epsilon) -f_a^0(s, \epsilon)\frac{\tilde{n_a}(s)}{n_{a0}(s)})- \gamma_{na}(s)f_a^0(s, \epsilon)\frac{\tilde{n_a}(s)}{n_{a0}(s)}
	\label{EqHeatingSource}
\end{equation}
is used, which allows to restore the density and temperature profiles on a long time scale, determined by the size of $\gamma_{ta}$ and $\gamma_{na}$ \citep{McMillan2008}. As in the case of the Krook operator, the time constants of this heating source term should be large enough as compared to the maximum growth rate and collision time in order not to perturbe the simulation results. 

The quantities $f_a^0(s, \epsilon)$, $\delta f_a(s, \epsilon)$ represent, respectively, the average values of the equilibrium and perturbed distribution functions  of species $a$   in a radial bin around the radial position $s$, and the kinetic energy $\epsilon$, of the marker. $n_{a0(s)}$ and $\tilde{n}_a(s)$ are the equilibrium and perturbation density around this position. Note that $n_{a0}(s)=\sum_{\epsilon} f_a^0(s, \epsilon)$ and $\tilde{n}_a(s)=\sum_{\epsilon}\delta f_a(s, \epsilon)$. As for the Krook operator, we will drop the species index from $\gamma_{ta}$, $\gamma_{na}$ because we only deal with a kinetic species, the main ions.

%***************************************************************************************
\subsection{Influence of NCSTs on linear growth rates in a tokamak}\label{secInflNCSTsLinGR}
\begin{figure}
	\centering
	\includegraphics[draft=false, trim=0 0 0 0, clip, height=3.35cm]{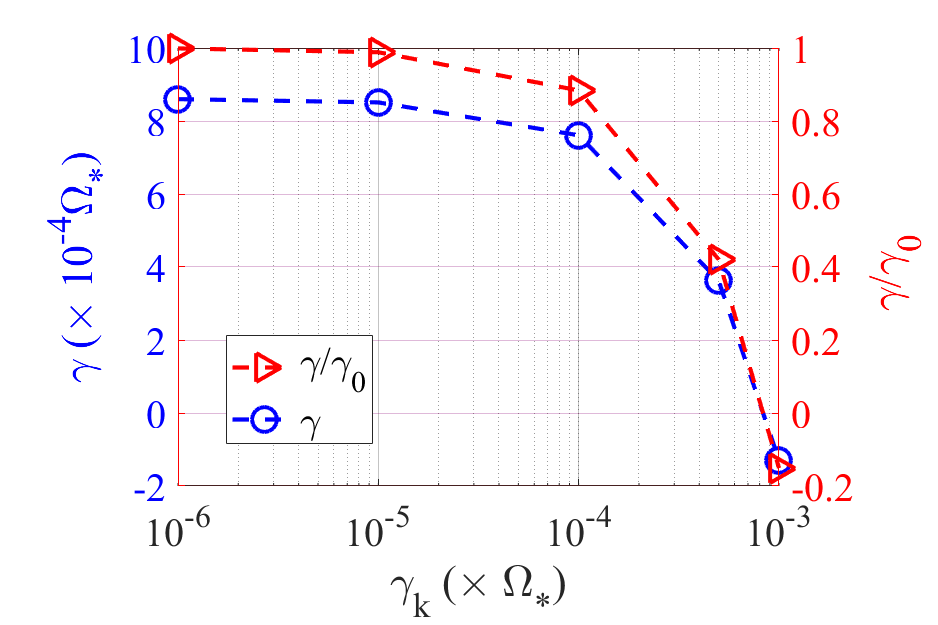}
	\caption{Linear growth rate of the most unstable mode in a set of linear simulations in the CBC MHD equilibrium using a Krook operator with different values of $\gamma_k$.}
	\label{figInfKrookLinCBC}
\end{figure}

%***************************************************************************************
Before going to the characterization of NCSTs in stellarators we first study how using these tools affects the growth rates of linearly unstable modes  in tokamak geometry. We use the same CBC MHD equilibrium and analytical density and temperature profiles given in Eq. (\ref{analProfilesnT}). 
First, we study the use of a Krook operator. We run a set of linear collisionless simulations using the same density and temperature  profiles from figure \ref{fig:profsNLBenchORB5EUT150M} and using a Krook operator with different values of the constant $\gamma _k = 10^{-6}, 10^{-5}, 10^{-4}, 5\times 10^{-4}, 10^{-3}$, in units of $\Omega_* ^{-1}$, with $\Omega_* = e B_* / m$ the ion cyclotron frequency,  and $B_*$ the average value of the magnetic field strength at the magnetic axis. The results of these simulations are shown in figure \ref{figInfKrookLinCBC}. The growth rate of the most unstable mode for the reference case, without Krook operator, is $\gamma _m=8.5 \times 10^{-4} \Omega_*$.
	
%*************************************************************************************
%*************************************************************************************
A shown in the figure, the Krook operator has a strong effect on the linear growth rate when $\gamma _k$ approaches the linear growth rate of the unstable modes. It can be inferred from these reults that keeping $\gamma _k < \gamma _m /10$ does not yield to alter the growth rate in more than 10\%.% We can deduce the common rule of thumb that keeping $\gamma _k < \gamma _m /10$ is safe, as it allows reducing the effect on the growth rate below a 10\%.

\begin{figure}
	\centering
	\includegraphics[draft=false, trim=20 0 30 0, clip, height=3.35cm]{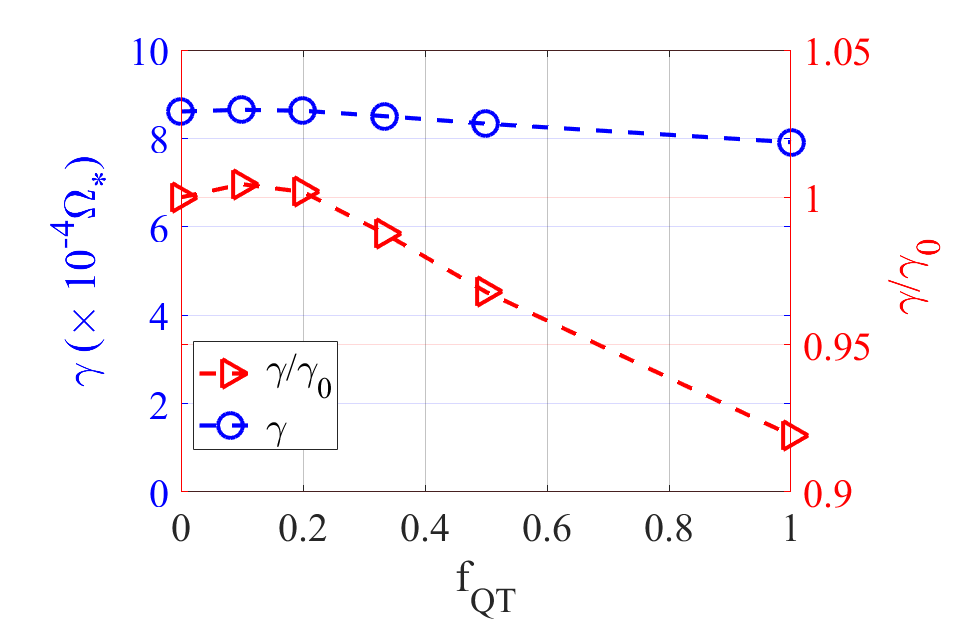}
	\includegraphics[draft=false, trim=50 0 30 0, clip, height=3.35cm]{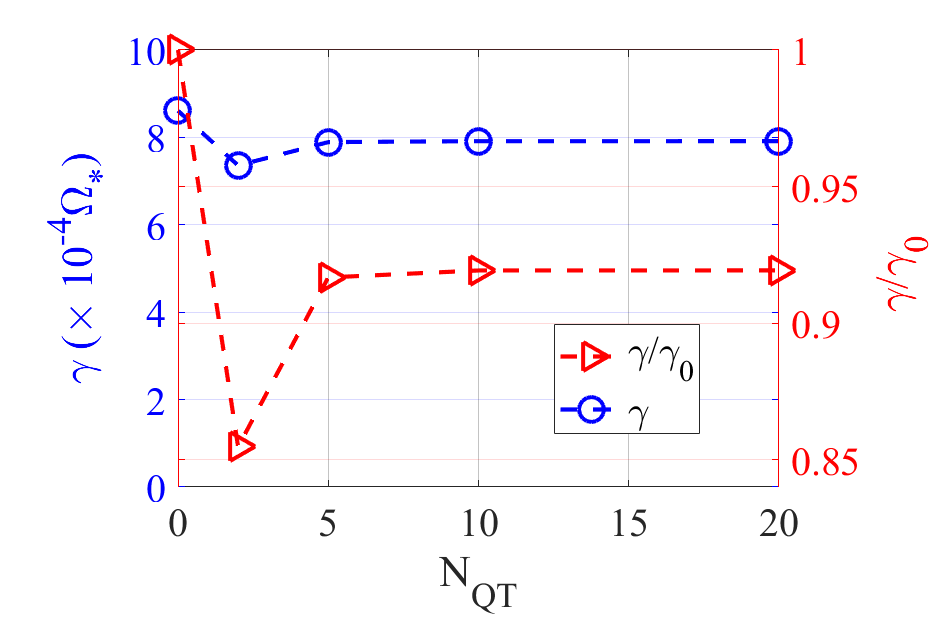}
	\includegraphics[draft=false, trim=50 0 10 0, clip, height=3.35cm]{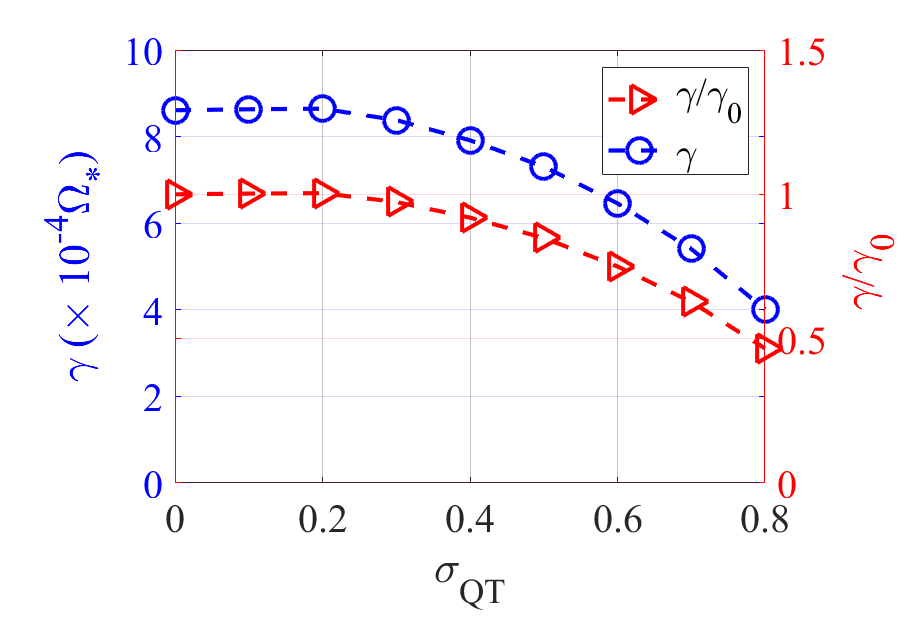}
	\caption{Influence of the weight smoothing on linear growth rate in the CBC equilibrium for different values of the frequency (left), minimum number of markers in each quad tree cell  (middle) and $\sigma_{QT}$ (right). The reference parameters, used when only one of them is changed, are $N_{QT}=10$, $\sigma_{QT}=0.4$ and $f_{QT}=1$.}
	\label{figInfQTLinCBC}
\end{figure}

In figure \ref{figInfQTLinCBC} the influence of using the weight smoothing is shown. 	
%*************************************************************************************
%*************************************************************************************
The parameter with a stronger influence on the growth rate is $\sigma_{QT}$, while the frequency and minimum number of markers per  bin have a weak influence. This is easily understood because $\sigma_{QT}$ enters the expressions (\ref{EqweightChange1}) and (\ref{EqweightChange2}) in an exponential function determining the distance in phase space at which the smoothing is effective.

\begin{figure}
	\centering
	\includegraphics[draft=false, trim=0 0 0 0, clip, height=3.35cm]{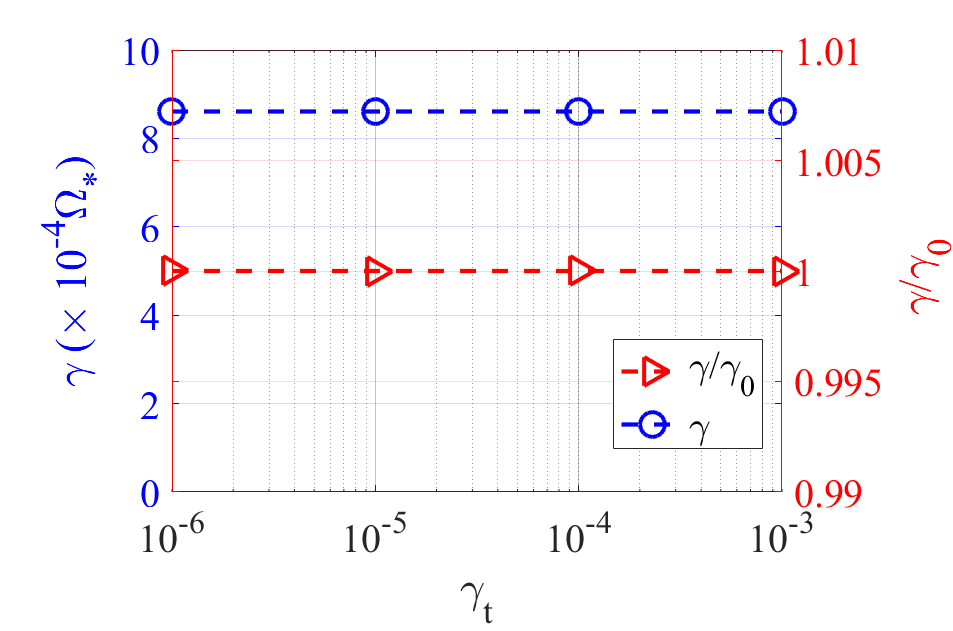}
	\includegraphics[draft=false, trim=0 0 0 0, clip, height=3.35cm]{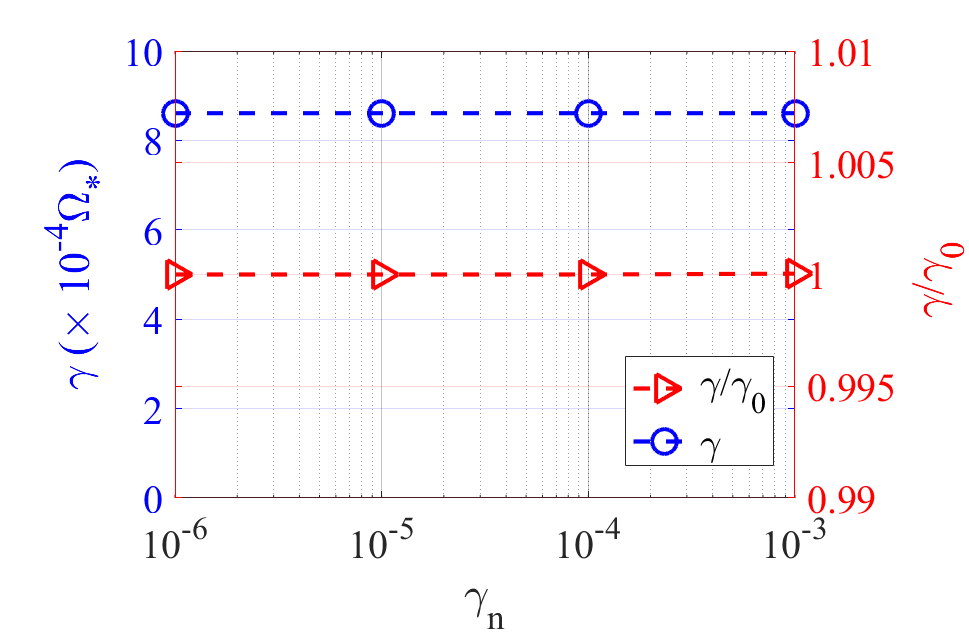}
	\caption{Influence of the heating source term on the linear growth rate in the CBC equilibrium for different values of $\gamma_t$ (left) and $\gamma_n$ (right).}
	\label{figInfHeatingTLinCBC}
\end{figure}

Finally, figure \ref{figInfHeatingTLinCBC} shows the results of a set of linear simulations with the same parameters in which only the heating source term is used with different values of $\gamma_t$ and $\gamma_n$. We show the results obtained by changing each parameter separately. It is clear that the heating source term has small influence on the linear growth rate even for values of $\gamma_t$ and $\gamma_n$ close to, or even slightly larger than, the maximum growth rate.
%*************************************************************************************
%*************************************************************************************

%***************************************************************************************
\subsection{Influence of NCSTs on linear properties in a stellarator configuration}
%***************************************************************************************
Now we turn to study the effect of the NCSTs in stellarator configurations.
We will use the standard magnetic configuration of W7-X (Ref\_11\_EIM) and model density and temperature profiles given by
\begin{eqnarray}
	X=X_* \rm{exp} \left\{ \frac{-\kappa _x}{1-\cosh^2(\frac{\rho_0}{\Delta _x})}  \left[\Delta _x \tanh(\frac{\rho-\rho_0}{\Delta _x}) - \cosh^2(\frac{\rho-\rho_0}{\Delta _x}) \right] \right\},
	\label{analProfilesnTW7X}
%	\label{analProfilesnW7X}\\
%	T_i=T_* \rm{exp} \left\{ \frac{-\kappa _{Ti}}{1-cosh^2(\frac{\rho_0}{\Delta _{Ti}})}  \left[\Delta _{Ti} tanh(\frac{\rho-\rho_0}{\Delta _{Ti}}) - cosh^2(\frac{\rho-\rho_0}{\Delta _{Ti}}) \right] \right\},
%	\label{analProfilesTW7X} 
\end{eqnarray}
% Eq. (\ref{analProfilesT}) such as those in figure \ref{fig:profsNLBenchORB5EUT150M}
where $X=\{n, T_e, T_i\}$ represents the profiles for density and electron and ion temperatures, and $X_*$ their values at the reference position $\rho_0$,
  with $\rho=r/a$, $\kappa_{Ti} =3$, $\kappa_{n} =1$, $\rho_0=0.5$, $\Delta_{Ti} = 0.1$, $\Delta_n=0.3$. The electron temperature profile is flat with $T_e=T_*=11.5 ~\rm{keV}$ and $n_*=1\times 10^{19} \rm{m}^{-3}$. The ion temperature profile has a maximum $T_i$ gradient at middle radius , $r/a=\rho_0=0.5$ . For this equilibrium we run a linear collisionless simulation with adiabatic electrons, without any source term or weight smoothing, in which the growth rate of the most unstable mode is $\gamma_{max}\sim 8.5\times 10^{-4}\Omega_*$. This is used as a reference simulation without using any NCST.
	
%***************************************************************************************
\subsection{Influence  of NCSTs on the linear growth rate}\label{secInflLinearGRINW7X}
%***************************************************************************************
%*************************************************************************************
\begin{figure}
	\centering
	\includegraphics[draft=false,  trim=25 0 40 10, clip, height=3.2cm]{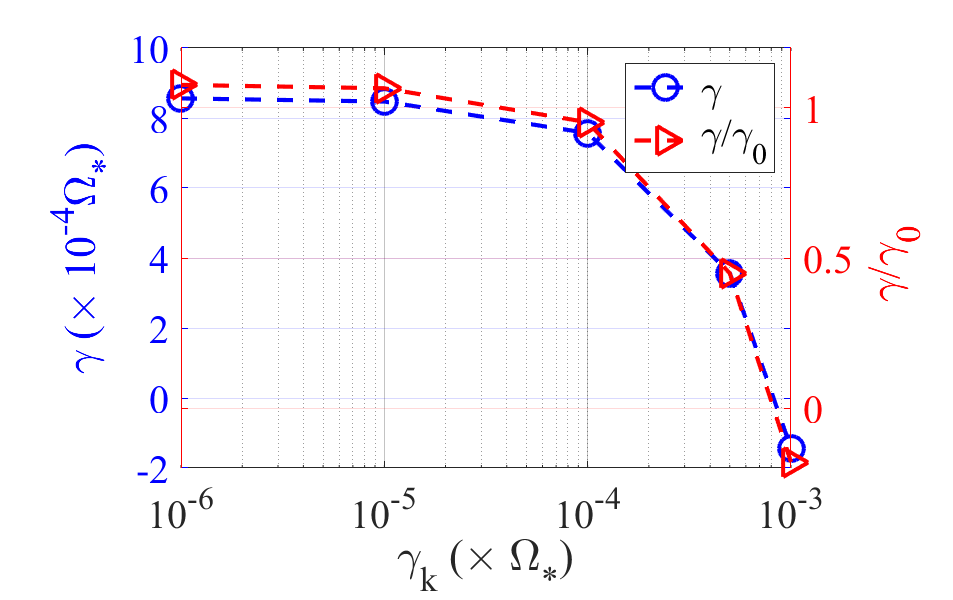}
	\includegraphics[draft=false,  trim=55 0 40 10, clip, height=3.2cm]{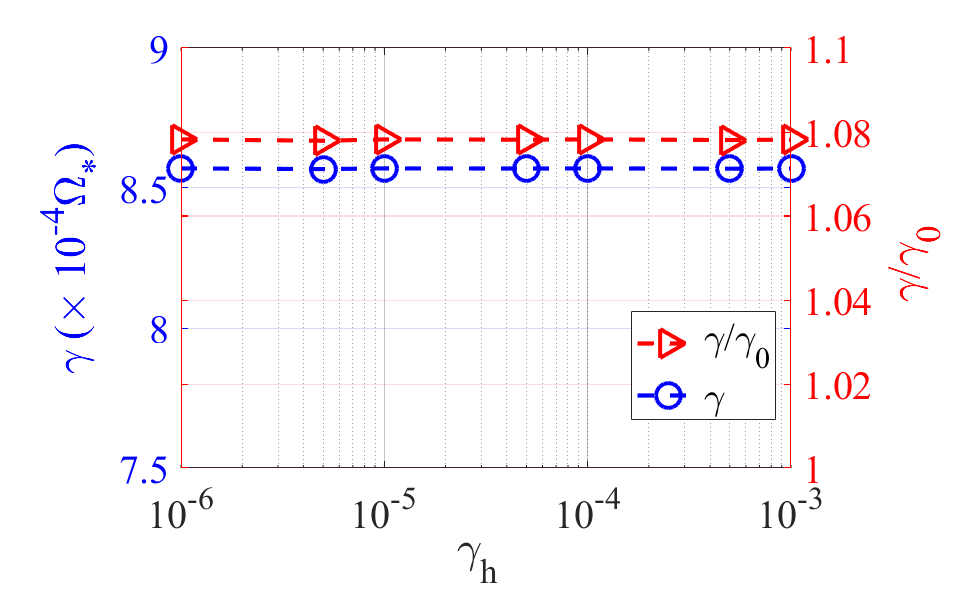}
	\includegraphics[draft=false,  trim=55 0 15 10, clip, height=3.2cm]{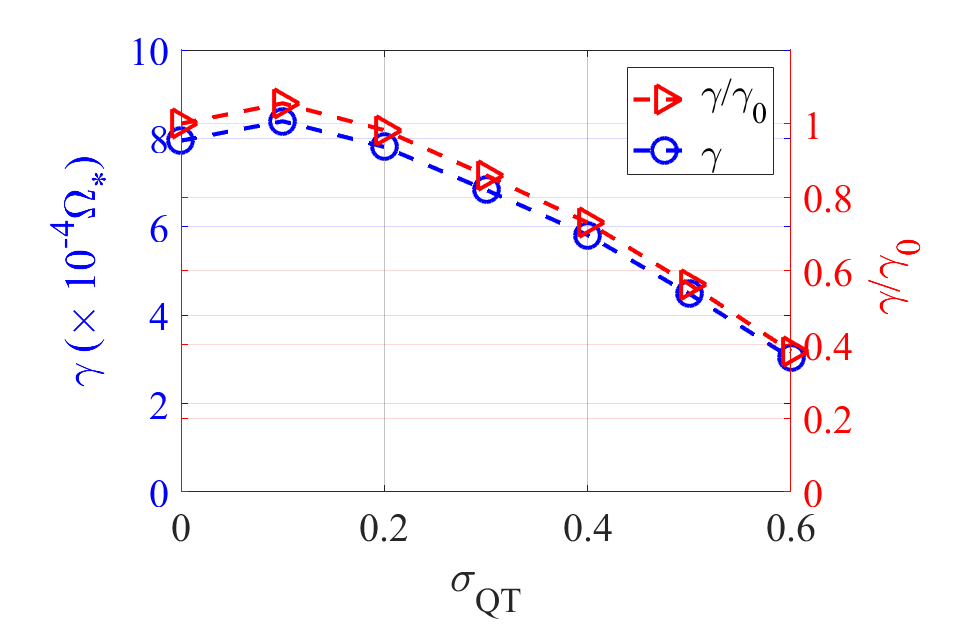}
	\caption{Influence of the Krook operator (left),  heating source, with equal values of $\gamma_t$ and $\gamma_n$, (middle) and weight smoothing for $N_{QT}=10, f_{QT}=\Omega_*$ (right) on the linear growth rate in a W7-X equilibrium (Ref\_11\_EIM) for different values of parameters. }
	\label{figInfQTLinW7X}
\end{figure}
%*************************************************************************************
In figure \ref{figInfQTLinW7X} the influence of Krook operator ($\gamma_k$), quad tree smoothing ($\sigma_{QT}$), and heating sources ($\gamma_t, \gamma_n$) on the linear growth rate of unstable modes is shown for a set of linear simulations in this W7-X configuration. As shown in the figure, and consistently with results from section \ref{secInflNCSTsLinGR}, the effect of the Krook operator on the linear growth rate is important when  $\gamma_k$ is close to the growth rate of the most unstable modes without any NCST used ($\gamma_{max} = 8.5 \times 10^{-4} \Omega_*$). The influence of the heating source is always smaller than 10\%, even for values of $\gamma_t$ and $\gamma_n$ similar or above the growth rate of the most unstable modes.

With respect to the weight smoothing, figure \ref{figInfQTLinW7X} shows the increasing effect of weight smoothing on the linear growth rate with the parameter $\sigma_{QT}$. It is clear that for  $\sigma_{QT}<0.2$ the effect on the growth rate is smaller than 10\%.
There is a slight dependency of the effect on the growth rate with the minimum number of markers in the phase space cell, $N_{QT}$, with an increasing effect as this number is reduced (not shown). This is interpreted as a result of the sharp cutoff associated to the exponential term in Eqs. (\ref{EqweightChange1}) and (\ref{EqweightChange2}). As the minimum number of markers in the bin is increased, the distance in phase space between markers whose weights are mixed is increased and then the exponential factor decreases strongly and the effect of smoothing is then reduced. On the contrary, for reduced numbers of markers the distance between them is reduced and the smoothing becomes more important. As the smoothing should be done only between markers with similar velocity components in order to preserve, as much as possible, the energy and momentum conservation, a rule of thumb can be derived that the number $N_{QT}$ should be reduced to the minimum for better conservation properties.  

%***************************************************************************************
\subsection{Influence of NCSTs on the linear evolution of zonal flows}\label{secInflNCSTLinZFs}
\begin{figure}
	\centering
	
	\includegraphics[draft=false,  trim=10 0 55 22, clip, height=3.15cm]{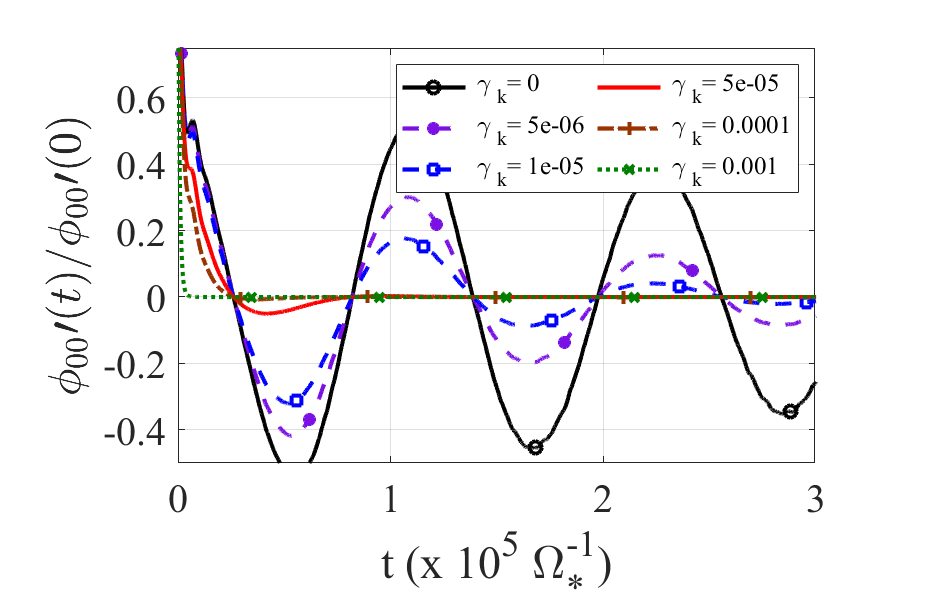}
	\includegraphics[draft=false,  trim=50 0 55 22, clip, height=3.15cm]{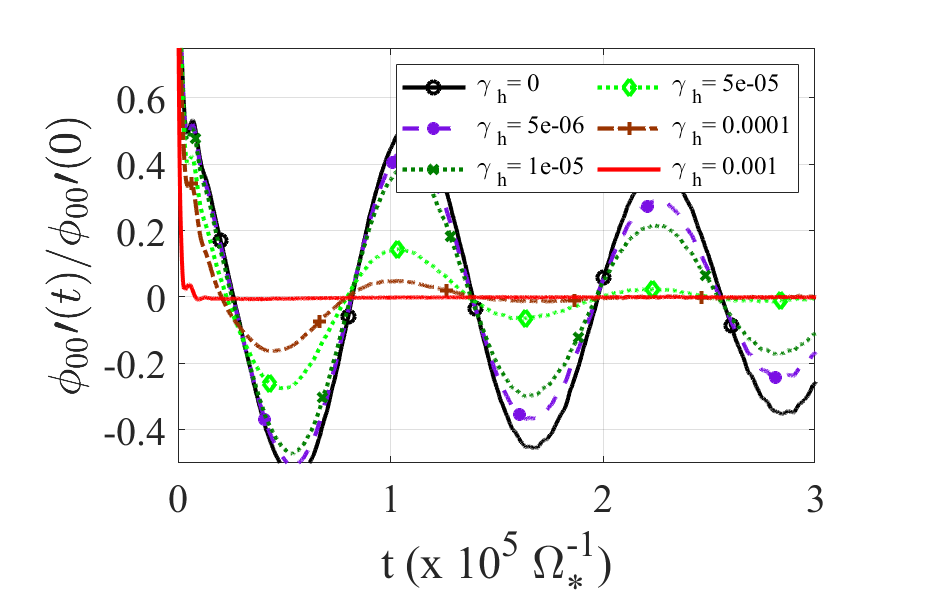}
	\includegraphics[draft=false,  trim=50 0 40 22, clip, height=3.15cm]{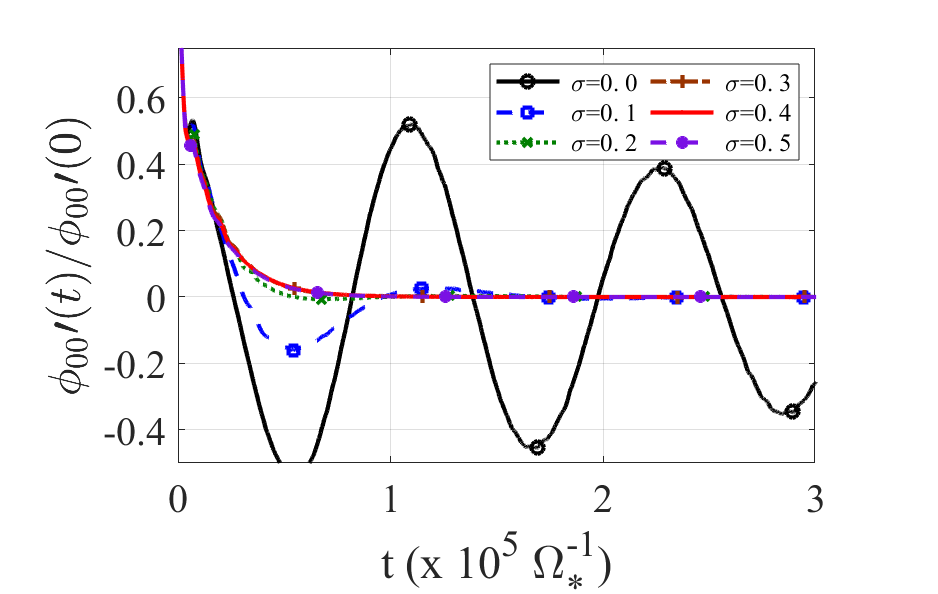}
	%\\
	\caption{Influence of the Krook operator (left),  heating source, with equal values of $\gamma_t$ and $\gamma_n$, (middle) and weight smoothing (right) on the linear evolution of the zonal potential component in a W7-X equilibrium for different values of parameters $\gamma_k$, $\gamma_h$ ($\gamma_t=\gamma_n=\gamma_h$) and $\sigma_{QT}$. In the right figure $N_{QT}=10, f_{QT}=\Omega_*$. }
	\label{figResNCSTOnZFW7X}
\end{figure}

%***************************************************************************************
%
In this section we study the influence of these tools on the linear evolution of zonal flows in a stellarator configuration, both the residual level \citep{Monreal2016} and the low frequency characteristic oscillation \citep{Mishchenko2008,Monreal2017}. 
We use the same standard configuration of W7-X as in Section \ref{secInflLinearGRINW7X}. In this case we use flat density and temperature profiles with $T_i=T_e=5 ~\rm{keV}$. We run linear simulations of zonal flow relaxation for a very small radial wavelength of the zonal flow. These simulations are run with adiabatic electrons and are started with a perturbation to the density with the form $\delta f \propto f_M \cos(k_s \pi s)$, with $k_s\rho_i<0.1$. Under these conditions the residual level is very small, in agreement with expectations \citep{Monreal2016}. First, we run a simulation with these parameters and without using any NCST, which we will use as a reference. In this simulation we observe a low frequency oscillation of the zonal potential with $\Omega_{ZF} = 8 \times 10^{-6} \Omega_*$.
Then, we run a set of linear simulations with the same setting but using the NCSTs with different parameters. The results are shown in figure \ref{figResNCSTOnZFW7X}.
%
%*************************************************************************************
%*************************************************************************************
The use of a Krook operator introduces a damping in the low frequency zonal flow oscillations. Its effect increases with the value of $\gamma_k$, and it has an important effect even on the short-time evolution of the ZF oscillation for values $\gamma_k > \Omega_{ZF}$.
The effect of using a heating source on the zonal flow oscillation is smaller than that of the Krook operator, with a smaller effect on the damping of ZF oscillations, even for values of $\gamma_h$ slightly above the reference zonal-flow frequency, $\gamma_h \sim \Omega_{ZF}$. It should be noted that the zonal flow oscillation frequency depends on the magnetic geometry and is typically much smaller than the typical growth rate of unstable modes \citep{Monreal2017}. %, in the range of $\gamma \sim 10^{-4}- 10^{-3}~\Omega_*$.
 In previous cases studied in W7-X, in section \ref{secInflLinearGRINW7X},  the maximum linear growth rate was $\gamma_{max} \sim 70 ~\Omega_{ZF}$ (at middle radius). Then, it can be expected than for a nonlinear simulation including realistic density and temperature profiles and zonal flow response, the parameter $\gamma_k$ of the Krook operator required for an effective noise control strongly affect the zonal flow evolution even at short times. The effect of heating sources can be expected to be slightly smaller, however.

\begin{figure}
	\centering
	\includegraphics[draft=false,  trim=22 0 45 22, clip, height=3.15cm]{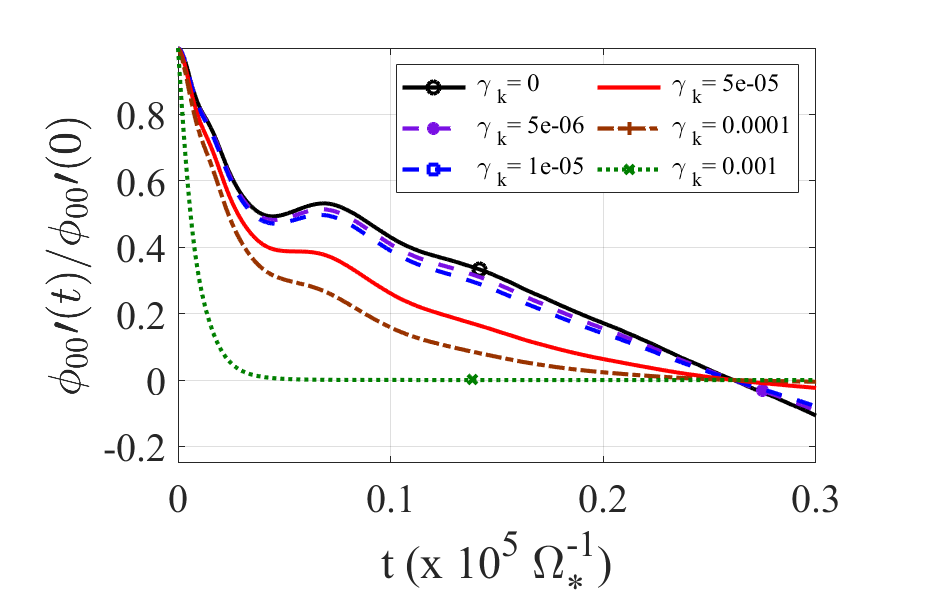}
	\includegraphics[draft=false,  trim=50 0 45 22, clip, height=3.15cm]{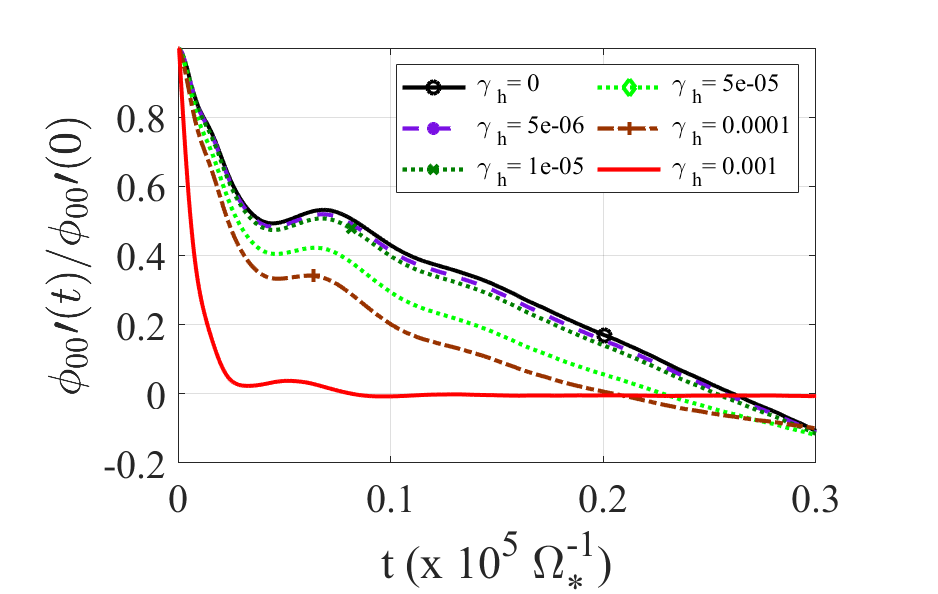}
	\includegraphics[draft=false,  trim=50 0 45 22, clip, height=3.15cm]{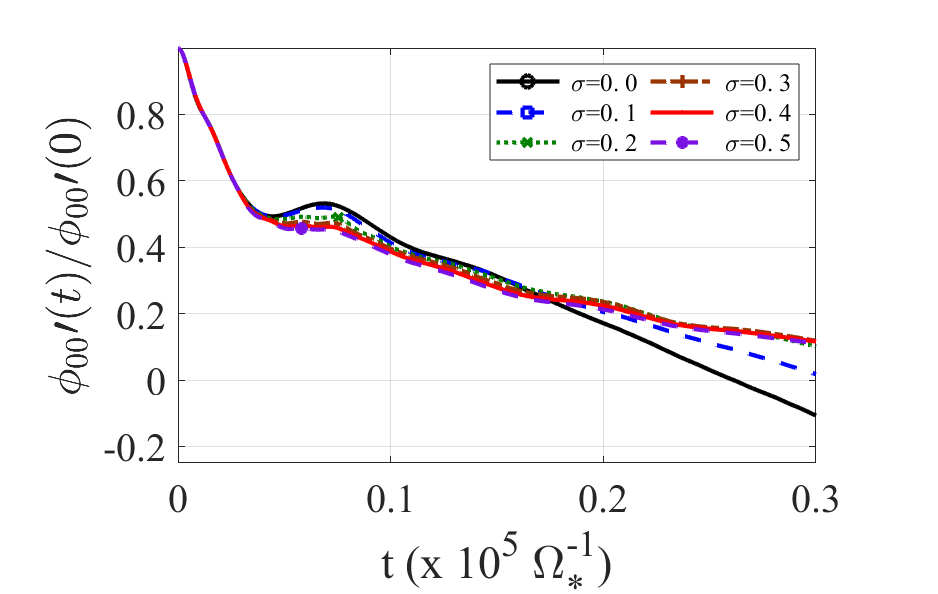}
	\caption{Influence of the Krook operator (left),  heating source, $\gamma_h$, (middle) and weight smoothing, $\sigma_{QT}$ (with $N_{QT}=10, f_{QT}=\Omega_*$) (right), on the linear evolution of the zonal potential component in W7-X equilibrium for different values of parameters. }
	\label{figInfQTLinW7XST}
\end{figure}

The effect of weight smoothing on the ZF oscillations is very strong for all values of $\sigma_{QT}$ studied, from $0.1$ to $0.5$ (always with $N_{QT}=10, f_{QT}=1 \Omega_*$). However, it is interesting noting here that the effect on the short times is very small for any value of $\sigma_{QT}$ as compared to the effect of both Krook operator and the heating source term. 
The effect of using these tools on the short time evolution of zonal flows is shown in more detail in figure \ref{figInfQTLinW7XST}.
It can be argued that the linear evolution of zonal flows for long times as compared to the characteristic turbulence times (eddy turn-over time, or inverse of the growth rate) should not be very relevant in the turbulence saturation. In this perspective, the weight smoothing could be considered as a method of noise control with small effect on the ZF, while the influence on ZFs at short times of heating sources, and particularly that of the Krook operator, is larger, always depending on the $\gamma_k$ and  $\gamma_h$ parameters values.

%*************************************************************************************
%*************************************************************************************

\begin{figure}
	\centering
	
	\includegraphics[draft=false,  trim=10 0 55 22, clip, height=3.15cm]{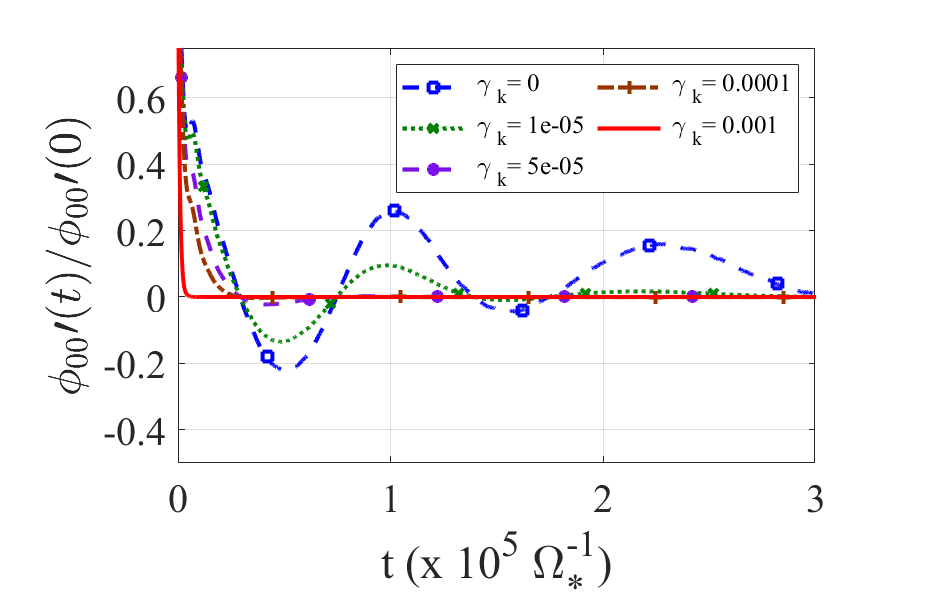}
	\includegraphics[draft=false,  trim=50 0 55 22, clip, height=3.15cm]{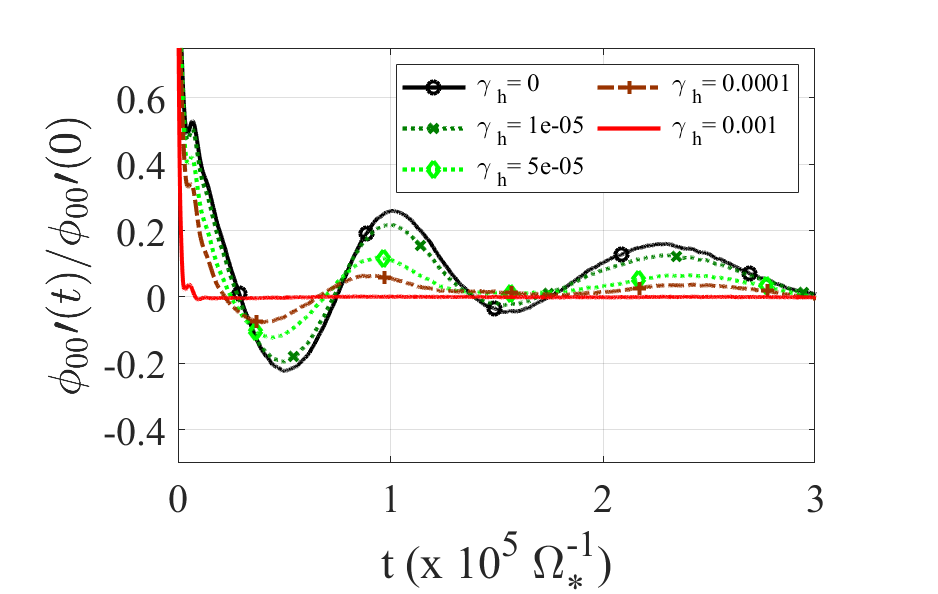}	
	\includegraphics[draft=false,  trim=50 0 55 22, clip, height=3.15cm]{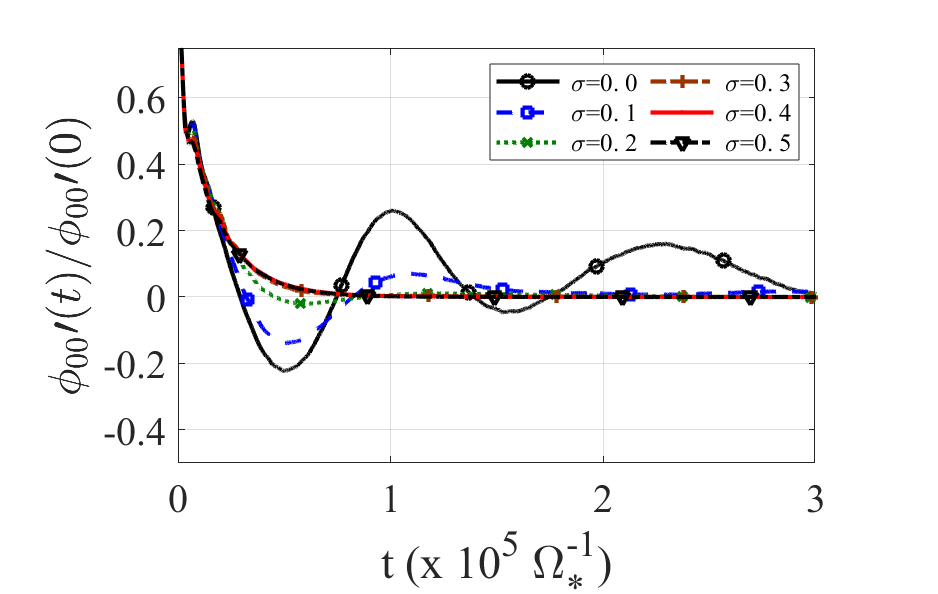}\\
	\caption{Influence of the Krook operator (left),  heating source, $\gamma_h$, (middle) and weight smoothing, $\sigma_{QT}$ (with $N_{QT}=10, f_{QT}=\Omega_*$) (right), on the linear evolution of the zonal potential component in W7-X equilibrium for different values of parameters including an electric field $\frac{d\phi}{dr}=r/a^2$. }
	\label{figInfQTLinW7XEr}
\end{figure}

\begin{figure}
	\centering

	\includegraphics[draft=false,  trim=15 0 50 22, clip, height=3.15cm]{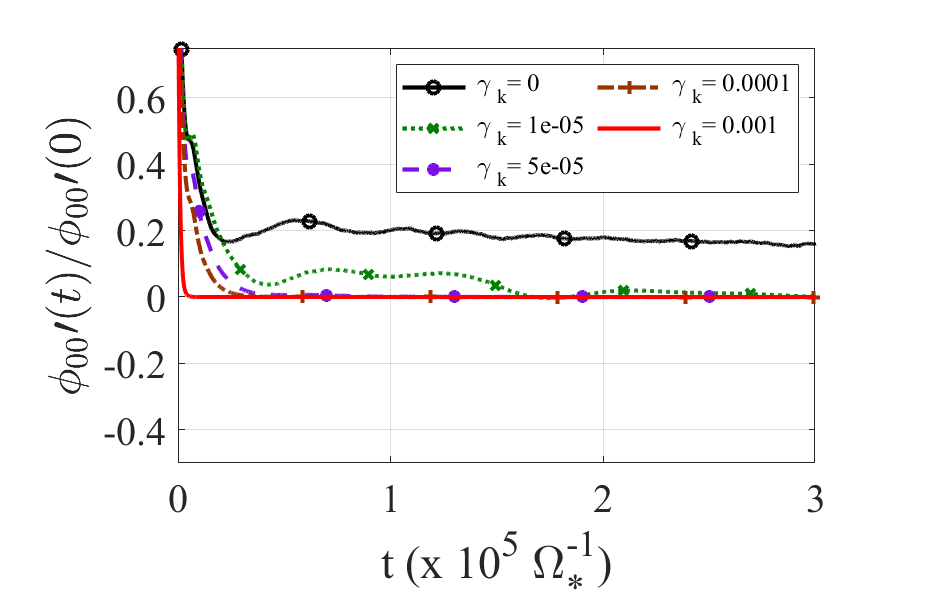}
	\includegraphics[draft=false,  trim=50 0 50 22, clip, height=3.15cm]{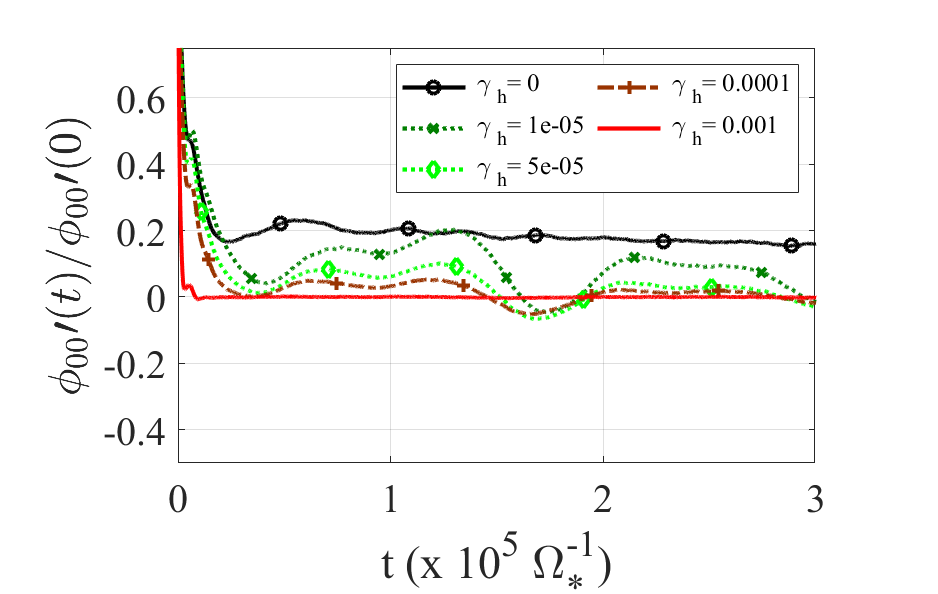}	
	\includegraphics[draft=false,  trim=50 0 50 22, clip, height=3.15cm]{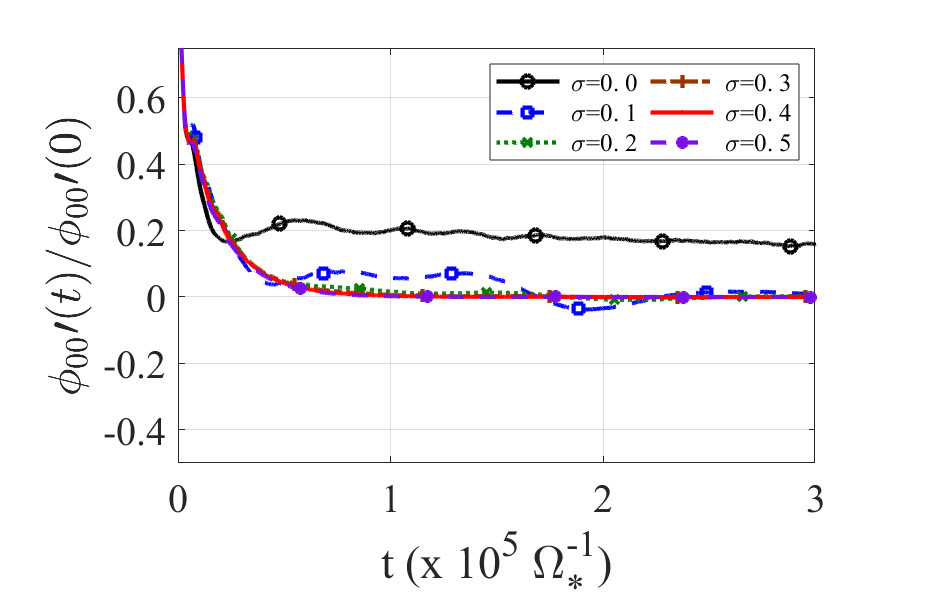}\\
	\caption{Influence of the Krook operator (left),  heating source, $\gamma_h$, (middle) and weight smoothing, $\sigma_{QT}$ (with $N_{QT}=10, f_{QT}=\Omega_*$) (right), on the linear evolution of the zonal potential component in a W7-X equilibrium for different values of parameters including an electric field  $\frac{d\phi}{dr}=2r/a^2$. }
	\label{figInfQTLinW7XEr2}
\end{figure}

A  background long-wavelength radial electric field is known to have an effect on both the residual zonal flow level \citep{Sugama2009b} and also on its oscillation frequency \citep{Mishchenko2012b}; then, it is worth checking the effect of using these NCST tools on the zonal flow evolution with a background electric field included in the simulation. The results of this test for two values of the background electric field are shown in figures \ref{figInfQTLinW7XEr} and \ref{figInfQTLinW7XEr2}. An electric field of the form $\phi'=\frac{d\phi}{ds}=0.5$ is used in the first case, shown in figure \ref{figInfQTLinW7XEr}, with $s=(r/a)^2$ being the radial coordinate, normalized toroidal flux.
%*************************************************************************************
%*************************************************************************************
By comparing figures \ref{figInfQTLinW7X} and \ref{figInfQTLinW7XEr} we can appreciate the effect of the electric field on the ZF oscillations: it introduces a damping on the oscillation. For this value of the radial electric field the change in this frequency and the residual ZF level is very small, however. The ZF oscillations are further damped when NCSTs are used, although the effect is slightly smaller than with no electric field (compare figure \ref{figInfQTLinW7XEr} and \ref{figInfQTLinW7X}). With respect to the influence at short times, the same conclusion obtained without electric field can be extracted in this case: quad tree smoothing is much less distorting at short times than using a Krook source term also when an electric field is included.

In figure \ref{figInfQTLinW7XEr2} we show the results for a series of simulations including a larger electric field, $\phi'=\frac{d\phi}{ds}=1$. In this case, the electric field introduces a finite residual level, an increase in the oscillation frequency and also a damping of these  oscillations.
%*************************************************************************************
%*************************************************************************************
As in previous cases, the use of NCSTs strongly affects the ZF relaxation with an increasing effect with parameters $\gamma_k$, $\gamma_t$, $\gamma_n$ and $\sigma$. The effect of the smoothing is very strong even for $\sigma=0.1$. However at short times the effect is not that large, and, anyway, much smaller than that of source terms when $\gamma_k$, $\gamma_t$ or $\gamma_n$ are above the ZF frequency.

%***************************************************************************************
\subsection{Influence of NCSTs on nonlinear simulations}
%***************************************************************************************
Once the effect of the NCSTs on the linear properties of unstable modes and zonal flows has been studied we turn to study their effect and effectiveness in controlling numerical noise in nonlinear simulations in a stellarator configuration. We will use the same standard configuration of W7-X and study nonlinear simulations using the same ideal profiles as in Section \ref{secInflLinearGRINW7X}, with $T_{*} = 11.5 ~\rm{keV}$. This unrealistic temperature value is used just to make the simulations less expensive in computational resources, which allows to run a set of them with different parameters at an affordable computational cost. We run simulations with adiabatic electrons. 
For these conditions, the growth rate of the most unstable mode, obtained from a reference linear simulation, is $\gamma_{max}=8.7\times 10^{-4}\Omega_*$, and the zonal flow oscillation frequency at middle radius is estimated as $\Omega_{ZF}\approx 1.2\times 10^{-5}\Omega_*$. This frequency is not calculated in a simulation, but estimated using that obtained in section \ref{secInflNCSTLinZFs} and the known scaling of this frequency with temperature \citep{Monreal2017,Sanchez2018} . A square Fourier filter is used in all these simulations in order to reduce the numerical noise. The squared filter is defined by its width in poloidal and toroidal directions of the Fourier spectrum. Only modes with $-m_f<m<m_f$ and $-n_f<n<n_f$, with $n_f=150$ and $m_f=170$ are kept in the simulation. In addition to the square filter, a diagonal filter, aligned with the field line, is superimposed, which suppresses all modes with $|m - n /\iotab  | > \Delta m$, with $\Delta m = 15$, because these modes have a too large parallel component. Note that only a period of the machine is simulated and, consequently, only modes with toroidal mode number being a multiple of the periodicity (five, in this case) are resolved.

In this study we will pay attention to several quantities that allow one to measure the quality of the simulation: the signal to noise ratio, the volume integrated zonal component and the non-zonal signal to signal ratio, 
which are defined as:
\begin{eqnarray}
		S2N & = & \frac{signal}{noise} = \frac{\sum_s^{n_s} \sum_{m,n \in F} |\phi_{m,n}(s)|^2}{\sum_s^{n_s} \sum_{m,n \notin F} |\phi_{m,n}(s)|^2}
		\label{eqS2NRatio}\\
		ZC  & = & \sum _s |\phi_{00}(s)|
		\label{eqZF}\\
		NZ2S & = & \frac{non-zonal \quad signal}{signal} = \frac{\sum_s^{n_s} \sum_{m,n \in F, n\neq0} |\phi_{m,n}(s)|^2}{\sum_s^{n_s} \sum_{m,n \in F} |\phi_{m,n}(s)|^2},
		\label{eqZFs2So}	
\end{eqnarray}
where $F$ represents the filter (intersection of the square and diagonal filters), $\phi_{mn}(s)$ is the $m, n$  Fourier component (in $\theta_*, \varphi$ coordinates, where $\varphi$ is the toroidal angle and $\theta_*$ is the straight-field-line poloidal coordinate) of the electrostatic potential, which is summed in a finite number of flux surfaces, ${n_s}$, at which the potential spectrum is evaluated.

\begin{figure}
	\centering
	\includegraphics[draft=false,  trim=30 0 17 5, clip, height=3.cm]{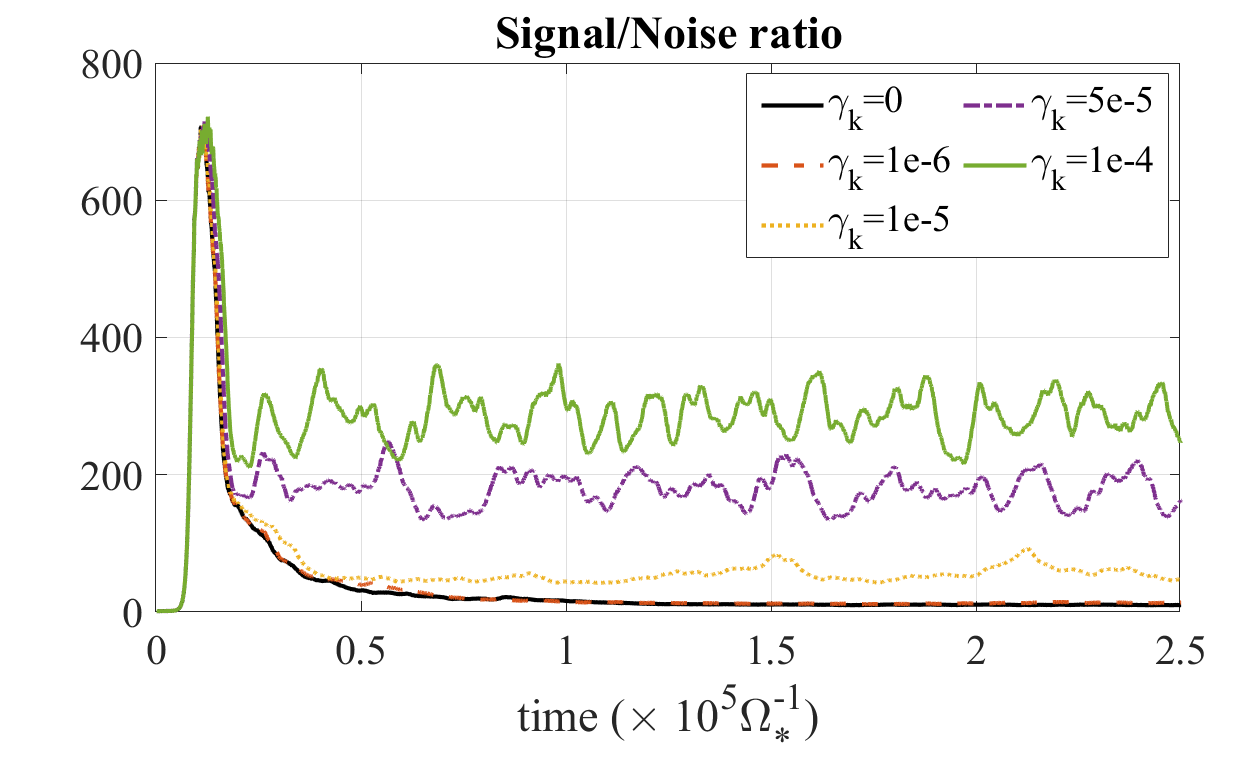}
	\includegraphics[draft=false,  trim=40 0 17 5, clip, height=3.cm]{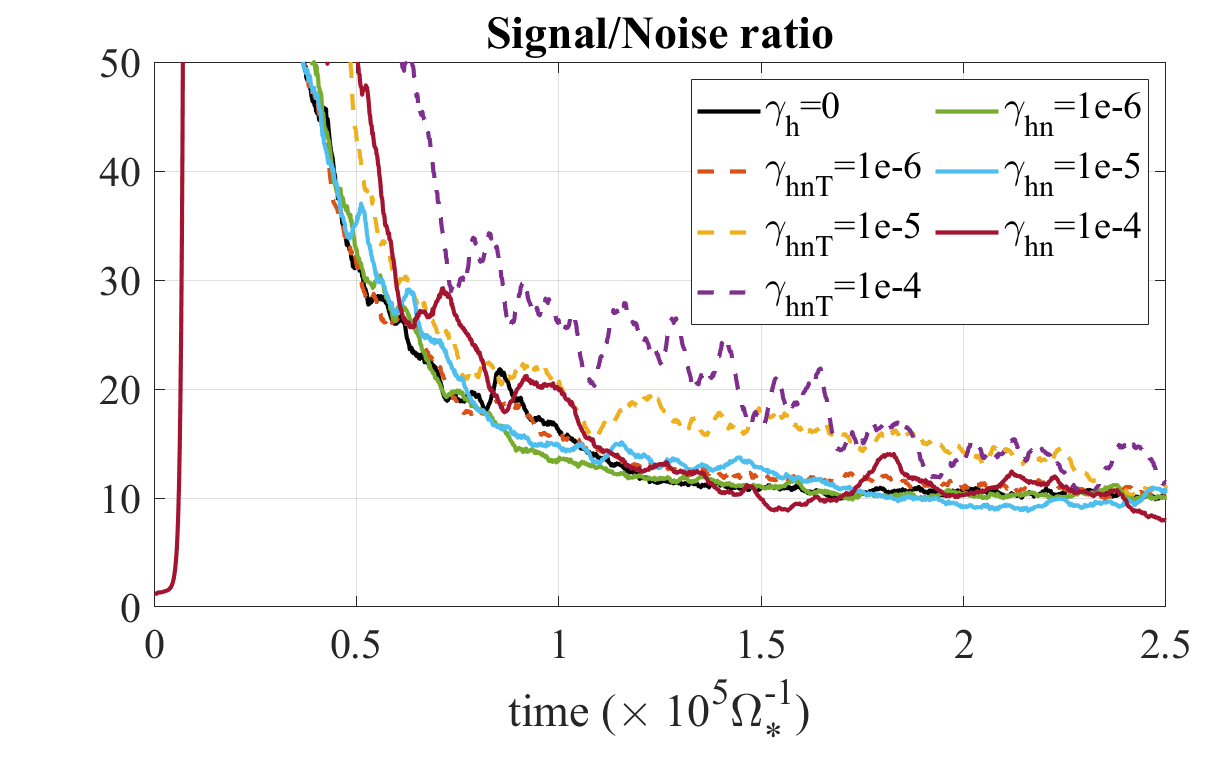}
	\includegraphics[draft=false,  trim=35 0 17 5, clip, height=3.cm]{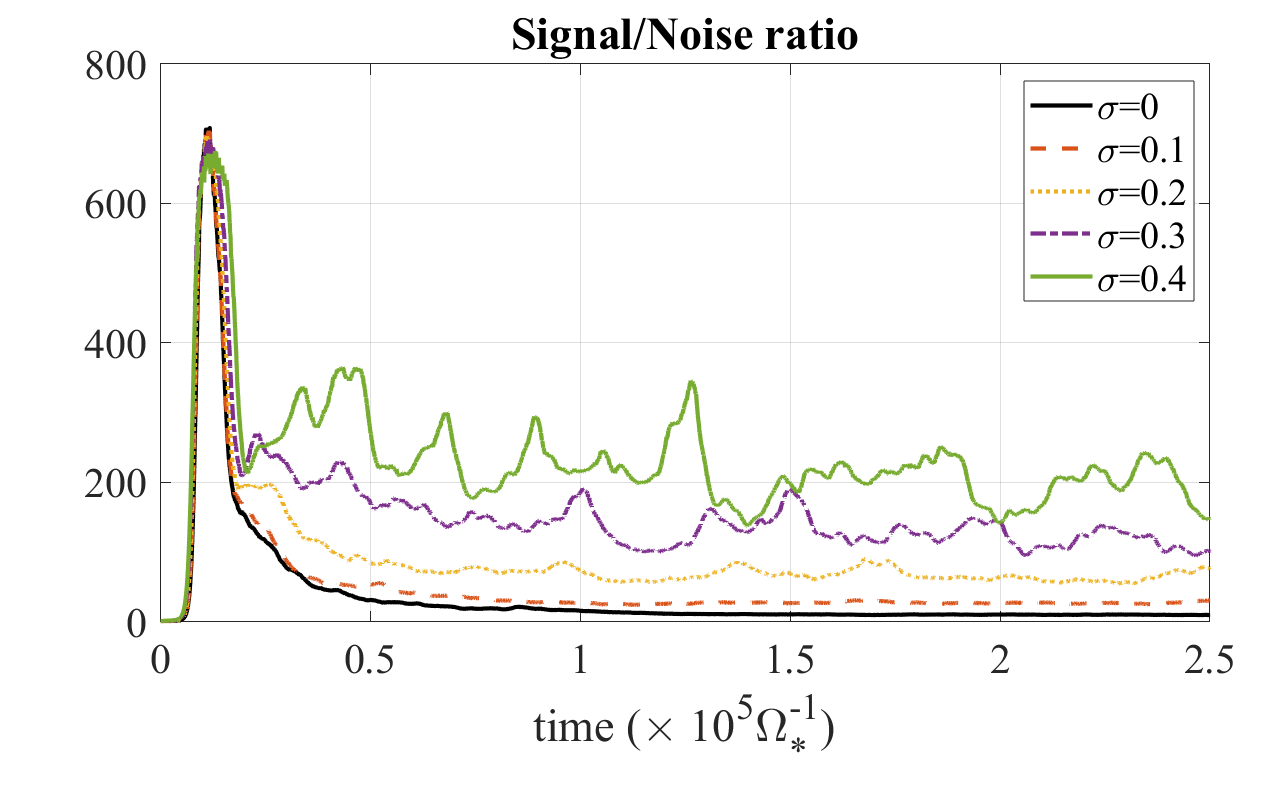}\\
	\caption{Influence of the Krook operator (left),  heating source, with equal values of $\gamma_t$ and $\gamma_n$, (middle) and weight smoothing for $N_{QT}=10, f_{QT}=\Omega_*$ (right) on the signal to noise ratio (defined in Eqs. \ref{eqS2NRatio}-\ref{eqZFs2So}) in a W7-X equilibrium for different values of parameters. }
	\label{figEvolutionS2NRatioW7X}
\end{figure}

\begin{figure}
	\centering
	\includegraphics[draft=false,  trim=5 0 20 5, clip, height=3.cm]{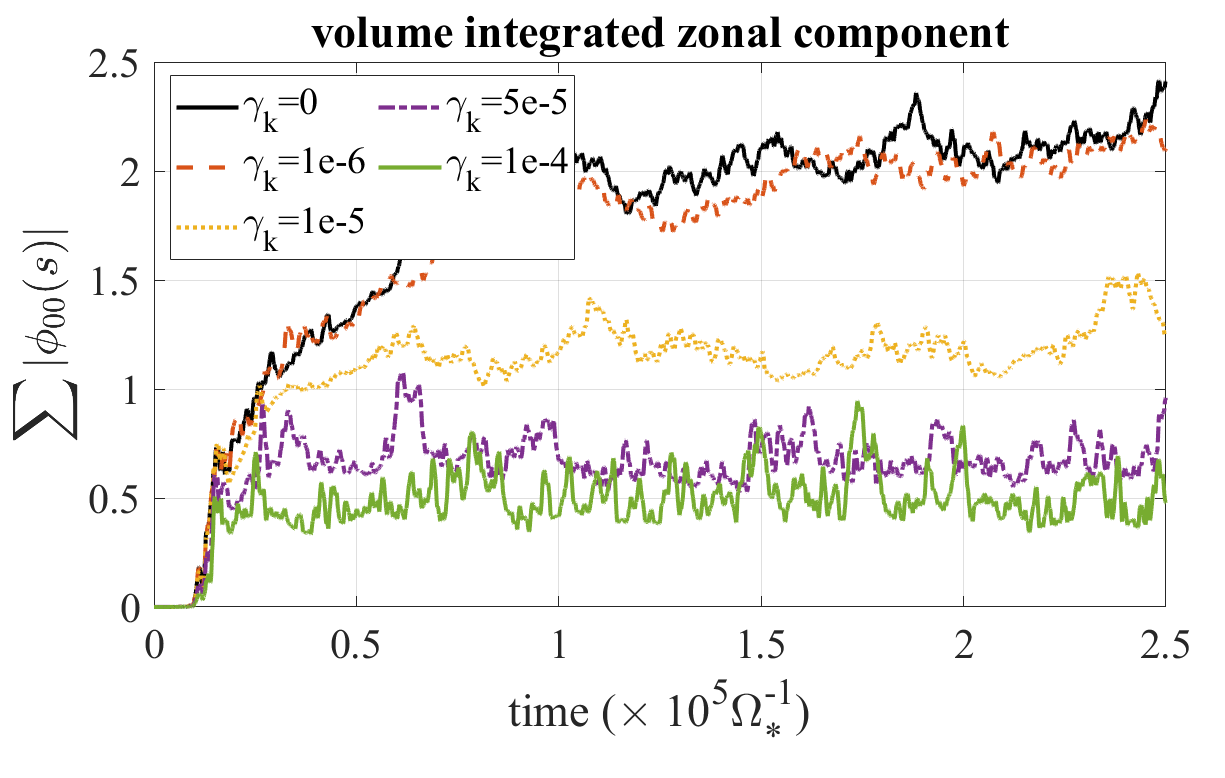}
	\includegraphics[draft=false,  trim=50 0 20 5, clip, height=3.cm]{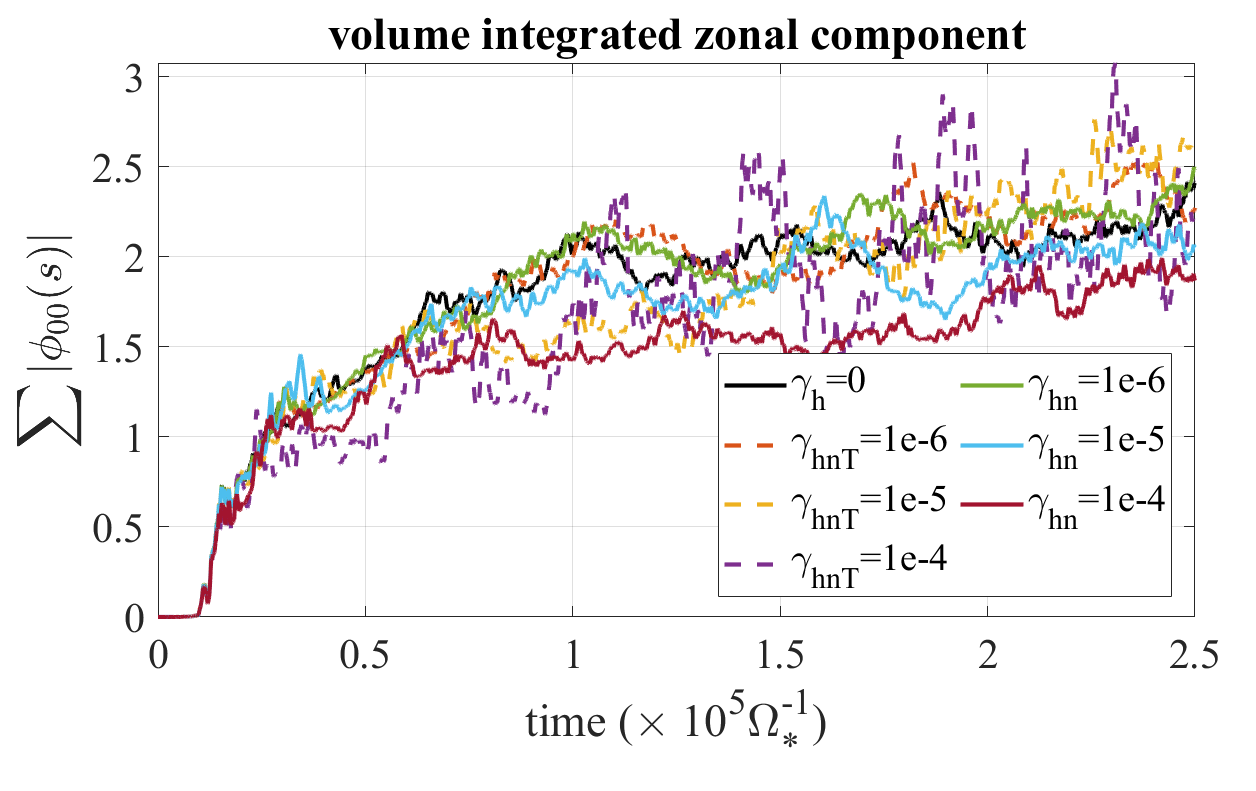}
	\includegraphics[draft=false,  trim=45 0 20 5, clip, height=3.cm]{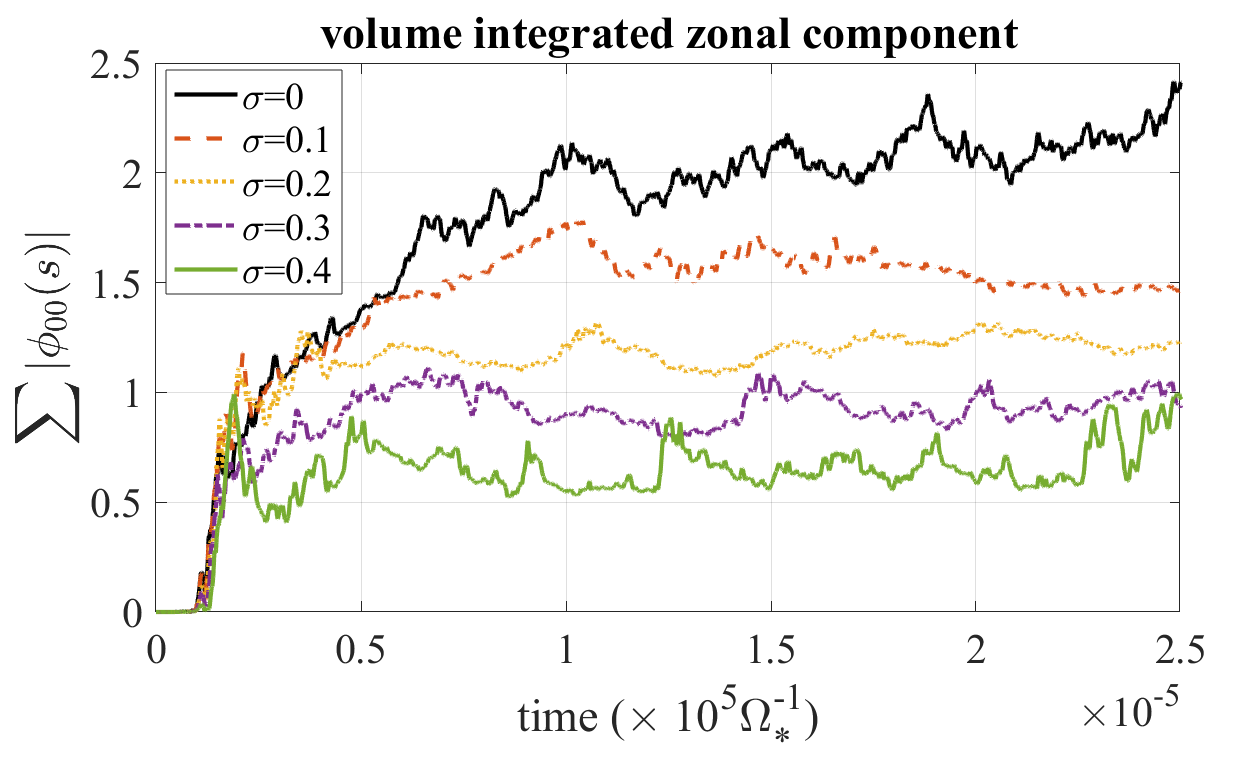}\\
	\caption{Influence of the Krook operator (left),  heating source, with equal values of $\gamma_t$ and $\gamma_n$, (middle) and weight smoothing (for $N_{QT}=10, f_{QT}=\Omega_*$) (right) on the zonal component in a W7-X equilibrium for different values of parameters. A volume integrated zonal component, as defined in Eq. (\ref{eqZF}), is used.}
	\label{figEvolutionZCW7X}
\end{figure}

  In addition, we will look at the volume integrated heat flux as a relevant physical quantity to study. In figure {\ref{figEvolutionS2NRatioW7X} the time evolution of the signal to noise ratio is shown for a set of simulations in which the NCSTs are used separately with different parameters $\gamma_k$, $\gamma_h$ and $\sigma_{QT}, N_{QT},  f_{QT}$. The reference case, without any NCST is shown in black color in figures \ref{figEvolutionS2NRatioW7X} and \ref{figEvolutionZCW7X}}.
%*************************************************************************************
%*************************************************************************************
It is clear from figure \ref{figEvolutionS2NRatioW7X} that the heating source term has a small effect on the $S2N$ ratio for any value of $\gamma_h$, while both Krook and quad tree smoothing have a significant effect. This means that a simulation in which only a source term is used, without including any  noise control,  will en up with the noise dominating the signal. The effect of both  Krook and QT is an improvement of the quality of the simulation as measured by the S2N ratio, and this improvement increases with he size of parameters $\gamma_k$ and $\sigma_{QT}$. Krook operator seems to be more efficient in increasing the S2N ratio. It is interesting to note that no convergence is found when these parameters are increased.
%*************************************************************************************
%*************************************************************************************

Another relevant quantity to look at is the zonal component (ZC).  In figure \ref{figEvolutionZCW7X} the volume integrated zonal component of the signal is shown versus time for several simulations with different parameters  $\gamma_k$, $\gamma_h$and $\sigma_{QT}$. The reference case, without any NCST is shown in black color in all plots.
It is clear that for noise uncontrolled simulations, the zonal component grows continuously in time without a saturation. This features was observed in previous non linear simulations with EUTERPE. This is related to the fact that the numerical noise directly contributes to the ZC. As for the S2N ratio, the heating sources have almost no effect on the ZC independently of the value of $\gamma_h$. However, both Krook and QT have a significant effect on ZC which increases as $\gamma_k$ and $\sigma_{QT}$ are increased, thus contributing to stabilize the ZC for long times. The effect of both tools is stronger over the ZC than over the S2N ratio and no clear saturation is observed in the range of parameters studied.

\begin{figure}
	\centering
	\includegraphics[draft=false,  trim=30 0 20 5, clip, height=3.cm]{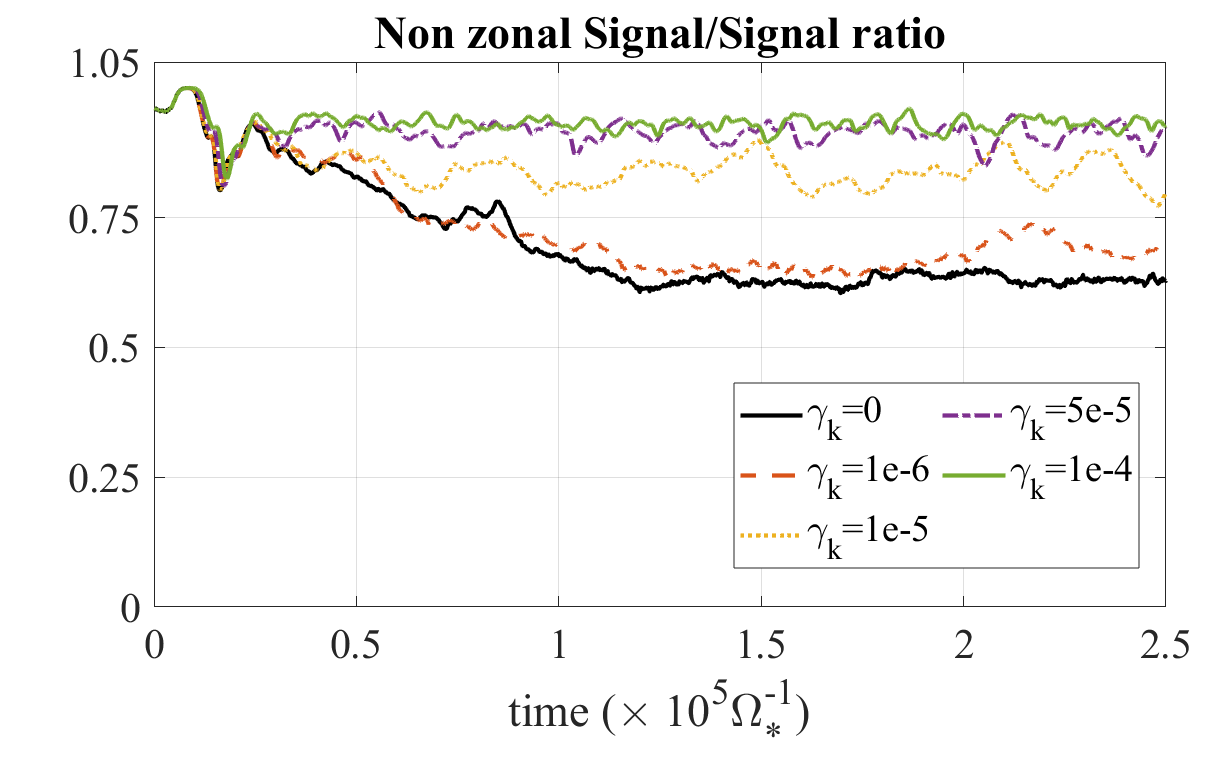}
	\includegraphics[draft=false,  trim=30 0 20 5, clip, height=3.cm]{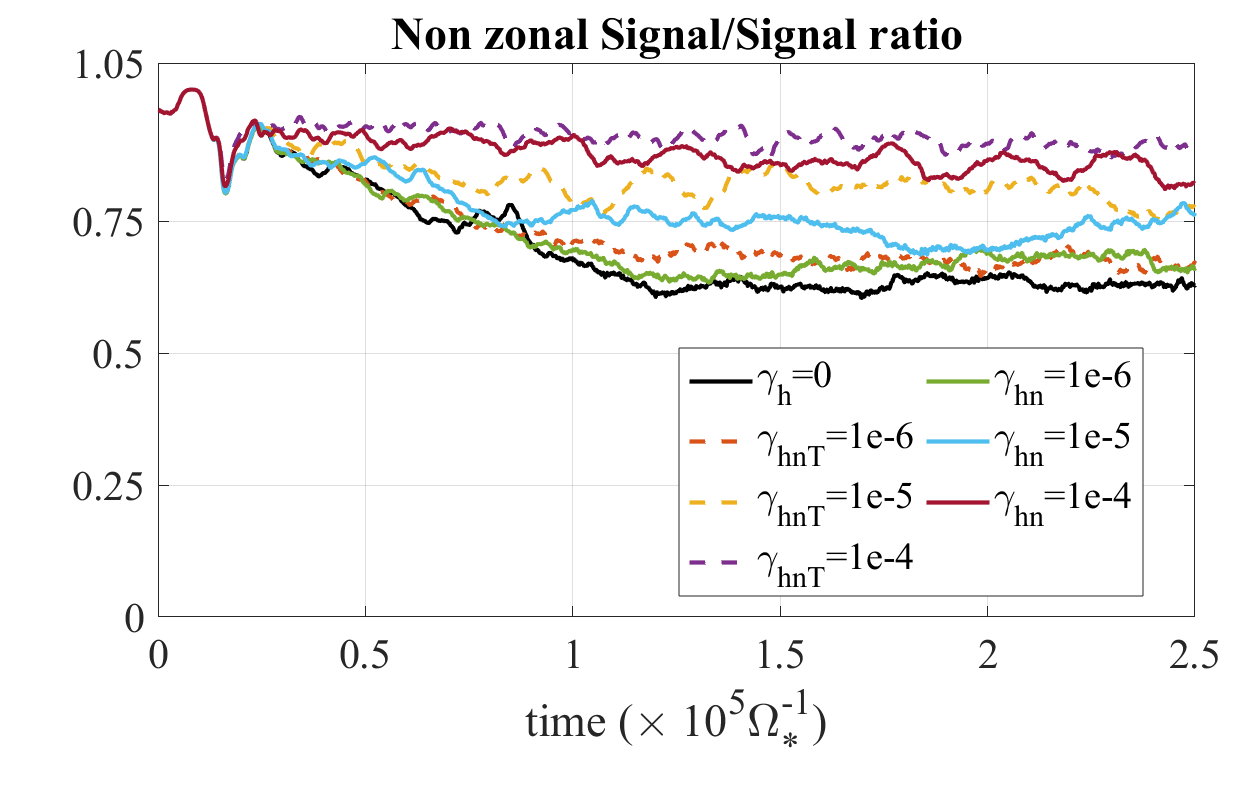}
	\includegraphics[draft=false,  trim=30 0 20 5, clip, height=3.cm]{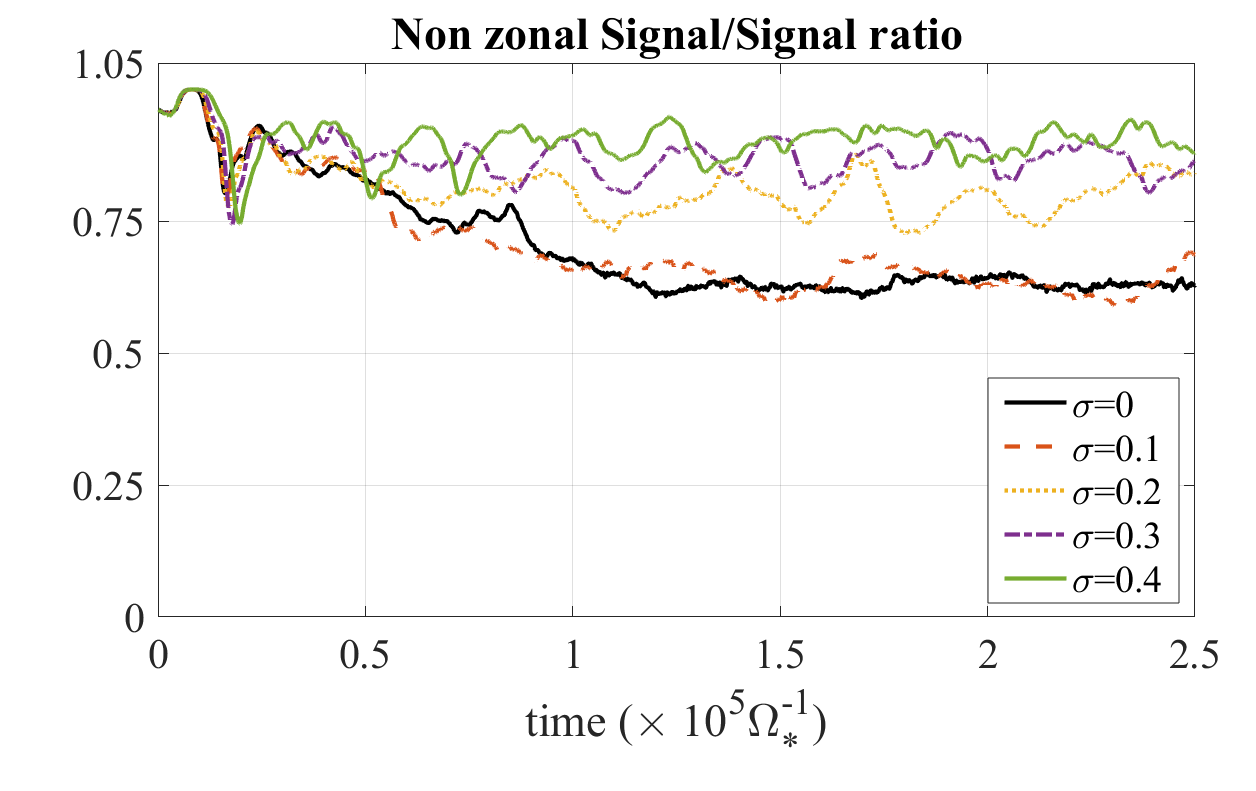}\\
	\caption{Influence of the Krook operator (left),  heating source, with equal values of $\gamma_t$ and $\gamma_n$, (middle) and weight smoothing for $N_{QT}=10, f_{QT}=\Omega_*$ (right) on the ratio of the non-zonal component to the signal in a W7-X equilibrium for different values of parameters. The ratio NZ2S is defined in Eq. (\ref{eqZFs2So}).}
	\label{figEvolutionNF2SRatioW7X}
\end{figure}

Now we look to the ratio of non-zonal component of the signal to the full signal, defined in Eq. (\ref{eqZFs2So}). 
This quantity is shown versus time in figure \ref{figEvolutionNF2SRatioW7X} for the same set of simulations used in previous figures.
%*************************************************************************************
%*************************************************************************************
In all cases, using a Krook operator, QT smoothing and heating sources, the NZ2S ratio increases as the parameters $\gamma_k$, $\gamma_h$ and $\sigma_{QT}$ are increased. As in figures \ref{figEvolutionS2NRatioW7X} and \ref{figEvolutionZCW7X}, the reference case, without any NCST is shown in black color. It is interesting to note in this figure that when no NCSTs is used, the NZ2S ratio saturates for times longer than $1.25 \times 10^{5} \Omega^{-1}$. %even for the reference case without any NCST. 
We interpret this as a consequence of the simultaneous increase of noise and zonal component for long times when no tools are used. This is a prove that a significant part of the zonal component of the signal can be driven by the noise instead of being related to the nonlinear interaction of physical modes.
When NCSTs are used the NZ2S ratio increases up to a saturation higher level, which is close to the level just after the nonlinear saturation (at $t\sim 0.25$).

Finally, we turn to look at the heat flux, as a physically relevant quantity.
%*************************************************************************************
\begin{figure}
	\centering
	\includegraphics[draft=false,  trim=20 0 18 5, clip, height=3.0cm]{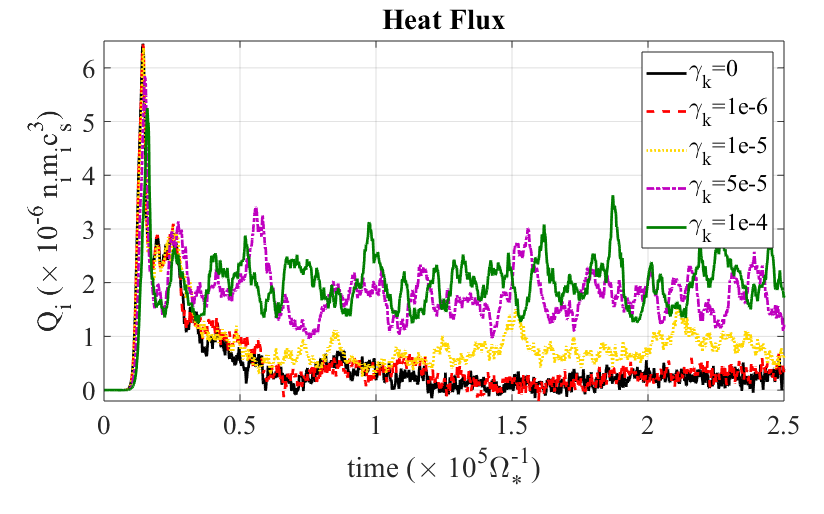}
	\includegraphics[draft=false,  trim=54 0 18 5, clip, height=3.0cm]{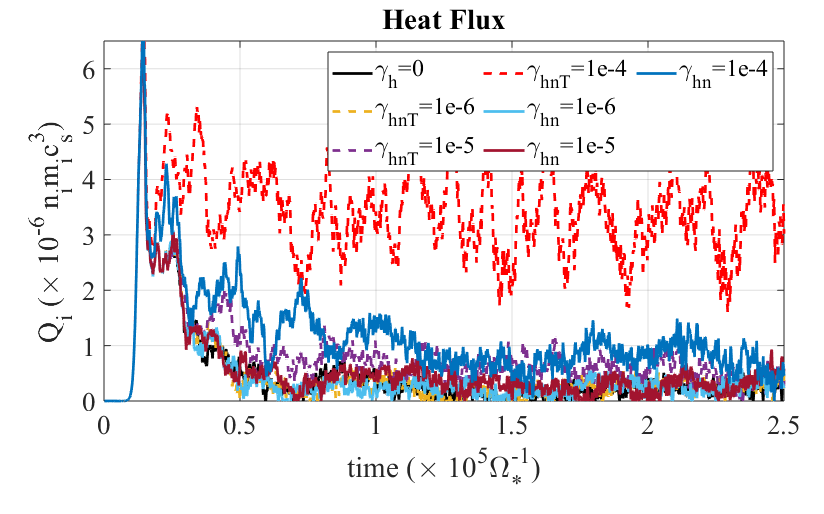}
	\includegraphics[draft=false,  trim=52 0 18 5, clip, height=3.0cm]{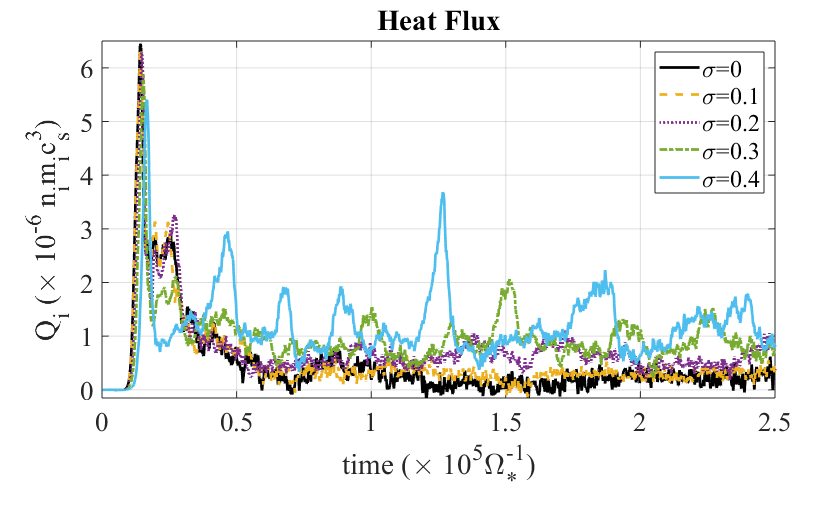}
	\\
	\caption{Influence of the Krook operator (left),  heating source, with equal values of $\gamma_t$ and $\gamma_n$, (middle) and weight smoothing for $N_{QT}=10, f_{QT}=\Omega_*$ (right) on the volume-integrated heat flux in a W7-X equilibrium for different values of parameters.}
	\label{figEvolutionHeatFluxoW7X}
\end{figure}
%*************************************************************************************
Figure \ref{figEvolutionHeatFluxoW7X} shows the time trace of the volume-integrated heat flux, as defined in Eq. (\ref{eqVolIntegHeatFluxDef}), for the same set of simulations studied before. As in previous figures, the reference case, without any NCST is shown in black color. 
It is clear that, if no NCSTs are used, the heat flux continuously decays after the initial overshot and the nonlinear saturation. This decay is the result of several effects. First, the instability has an effect on the profiles, which tends to relax the temperature gradient which drives the instability (ITG). On the other hand, the increase in the relative contribution of the uncorrelated numerical noise has also an effect on reducing the heat flux, which is produced by correlated fluctuations of density and velocity.

 When a Krook operator is included, the heat flux is stabilized to a value that increases with $\gamma_k$ and shows signs of some saturation for large-enough values  $\gamma_k>5\times 10^{-5}\Omega_*$. Note that these are values close to $\gamma_{max} /10 $ as shown in section \ref{secInflNCSTsLinGR}. As discussed there, for values $\gamma_k > \gamma_{max} /10 $ a significant effect of the Krook operator on the linear growth rate of unstable modes can be expected, and the effect on the linear evolution of ZFs is strong even for smaller values of $\gamma_k$.
With respect to the density and heating sources, we show in the figure the effect of switching on both density and heating sources separately. We can see that the  heating is more effective in modifying the heat flux. This can be expected as it is the temperature gradient that drives the underlying instability (ITG). Then, restoring the ion temperature profile has a large effect on the heat flux. The effect of the density source on the heat flux is very small, and it is small in any case except that $\gamma_h$ is large as compared to the growth rate of unstable modes.
Finally, the effect of QT smoothing on the heat flux is larger than that of density and heating sources but smaller than that of Krook operator. A clear stabilizing effect is observed on the heat flux signal for values $\sigma_{QT} > 0.2$ with no significant effect on the heat flux level, which is a kind of saturation. The saturation values of the heat flux are significantly smaller for the case of QT than for the case of using Krook operator.

After the comparison of several tools and their effect on the quality and physical results of the simulation, a short summarizing discussion is in order. As previously shown, the heating sources have a small effect on the simulation quality. Then, we can not rely on them for reducing the numerical noise. However, they can be used to stabilize the heat flux, and derived physical quantities, after the nonlinear saturation. It should be taken into account that the parameters $\gamma_h$ should be smaller than the relevant growth rate but, on the other hand, sufficiently large to sustain the equilibrium profiles without affecting the heat flux significantly.
The Krook operator appears as an interesting tool for both the noise control and the heat flux stabilization. However, the strong dependency of physical quantities (growth rate and heat flux) with the parameters used makes it less adequate, particularly in the case of collisional simulations in which a Krook operator with values of $\gamma_k$ large enough as to contribute to the noise control can disturb significantly the effect of collisions.
Another important drawback of using a Krook operator is the effect  on the linear evolution of zonal flows, which becomes important even at short times when $	\gamma_k$ is large enough as to produce benefitial effects on the S2N ratio, as shown in figure \ref{figInfQTLinW7XST}. An energy and momentum conserving scheme as that implemented in ORRB5 could mitigate this negative feature, but implementing this for stellarator geometry is not as clear as in tokamak.
Related the QT smoothing, it appears as an efficient tool for controlling the numerical noise and the secular zonal component which is driven by it. In addition, and similar to the Krook operator, QT smoothing has a stabilizing effect on the heat flux.  As a positive point we can remark that it has a very small effect on the linear ZF evolution at short times, which is expected to be the most relevant part for the nonlinear saturation of linearly unstable modes. As a drawback of this tool we can remark the lack of energy and momentum conservation, as in the process of smoothing the markers weights both quantities are not conserved. An energy and momentum conserving implementation, proposed in \citep{Donnel2019} has been tested in EUTERPE and the result is not very promising. When energy conservation is activated, then the effect of the QT smoothing is reduced up to almost not affecting the simulation, and its positive effect over the noise is lost. A partially conserving scheme is now under consideration.
After these tests of the three NCS tools, it appears that for realistic cases, including collisions, a combination of heating sources with appropriate (minimum) parameters to sustain the profiles and a soft weight smoothing can be more adequate than using a Krook operator.

Before going to a more practical case, we next study how the heat flux value is affected when the radial limits of the physical domain of simulation are changed. This is important from a practical point of view, as reducing the radial extension of the computational domain can contribute to reduce the computational resources required for nonlinear global simulations, which in general, are very expensive. 
	
\subsection{Radially limited simulations}

\begin{figure}
	\centering
	\includegraphics[draft=false,  trim=0 0 0 25, clip, height=4.25cm]{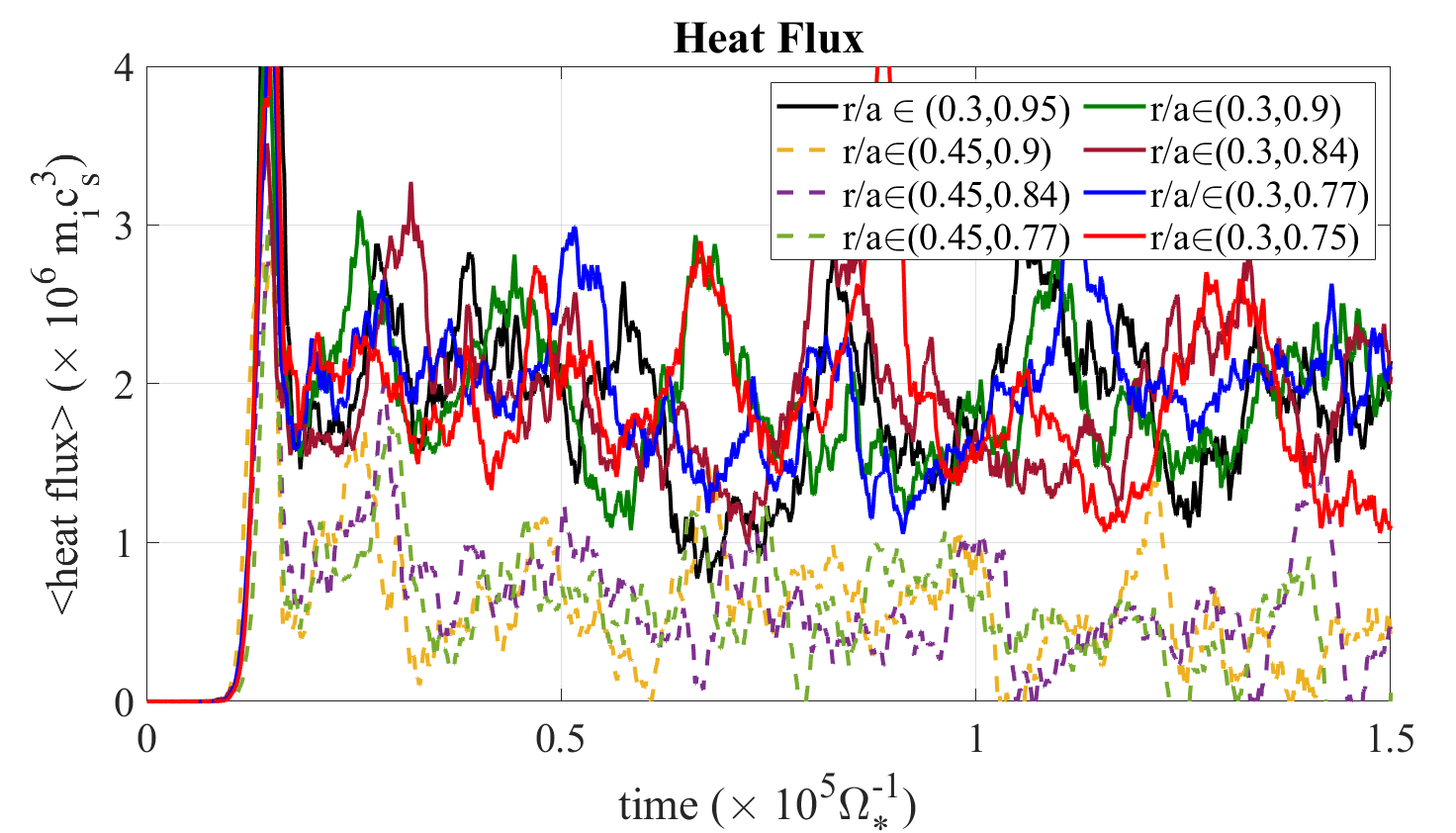}
	\\
	\caption{Influence of the radial limits of the computational domain on the volume-integrated heat flux in a W7-X equilibrium for different locations of the radial boundaries. In all cases a Krook operator is used with $\gamma _k = 5\times 10^{-5} \Omega_*$.}
	\label{figEvolutionHeatFluxRDW7X}
\end{figure}

We have run a set of simulations with the same equilibrium from W7-X and profiles previously used in which the limits of the radial domain are changed. In all the simulations the characteristic scale lengths for density and ion temperature are $a/L_{Ti}=3$ and $a/L_n=1$, the simulations are carried out with adiabatic electrons, using the long wavelength approximation and  a Krook operator, with $\gamma _k = 5\times 10^{-5} \Omega_*$, is used to stabilize the simulations. 
The volume-integrated heat flux is shown versus time for this set of simulations in figure \ref{figEvolutionHeatFluxRDW7X}.

%*************************************************************************************
%*************************************************************************************
The figure clearly shows that the heat flux level is not very sensitive to the location of the radial boundaries as far as the region of most unstable modes, which in this case are concentrated around middle radius, is captured in the simulation (solid time traces in the figure). 
However, the heat flux level is significantly reduced when the inner radial limit  is increased above $r/a=0.4$, thus entering the location of unstable modes, and affecting them (dashed lines in the figure).
In these simulations the instability parameter, $\eta_i=L_n/L_{Ti}$, is larger than 1 in the radial region $0.37 < r/a< 0.63$.

In practical cases, the criterion for limiting the radial domain may not be as clear as in this ideal case, as many different modes with different growth rates can coexist in different radial regions of the plasma, largely  depending on the density and temperature profiles. However, this exercise gives us confidence on the limited effect of the special boundary region on the heat flux produced by modes located inside the simulation volume.

%***************************************************************************************
\section{Application to a realistic W7-X case}\label{secApplW7XReal}
%***************************************************************************************
Now we turn to study nonlinear stellarator simulations in a more realistic scenario. The purpose of this section is to demonstrate the use of the NCSTs in a stellarator configuration using relevant experimental plasma parameters of density and temperature and including collisions. 
 For this purpose we have selected an experimental program carried out in the standard magnetic configuration of W7-X (configuration Ref\_170\_EIM).  
 We choose  the W7-X experimental program 20181016\_37, which has been extensively studied in \citep{Bozhenkov2020}, is well diagnosed and was used for power balance analysis. In this program, ECRH heating in the X2-mode was used, reaching an injected power of 5 MW.
Much attention has been put in this program because cryogenic pellets were injected, thus increasing both the central density above $8 \times 10^{19} \rm{m}^{-3}$ and peak ion temperature of $3~\rm{keV}$, and reaching record diamagnetic stored energy larger than $1.1 ~\rm{MJ}$. 
The focus of previous gyrokinetic analysis \citep{Xanthopoulos2020}
% and similar programs \citep{Stechow2020} 
has mainly been put on the external region of the plasma, $r/a>0.5$, where significant changes in profiles are produced in the post-pellet phase which lead to an increased confinement \citep{Bozhenkov2020}. However, we are now interested on testing the use of the NCSTs using experimentally realistic parameters with affordable resources. Computational resources required can be huge for the ion temperatures   typical from the plasma edge in W7-X, because the  wavenumber of the unstable modes increases as the temperature decreases and then, the spatial resolution has to be increased consequently. In addition, a wide spectrum of unstable modes (in terms of normalized wavenumber, $k_\bot\rho$) has been found in previous linear simulations in the outer region of W7-X (to be published), which further increases the computational resources required for simulating the plasma edge. This is the reason why we focus here in the innermost part of the plasma ($r/a<0.5$), where the ion temperature is above $1~\rm{keV}$. In this radial region of the plasma, during the post-pellet phase (seconds 1-2), the density profile is peaked and the ion temperature profile is quite flat, thus showing no ITG instability, which contributes to reduce turbulent transport in the core \citep{Stechow2020}. The stability of these profiles to ITG modes was confirmed with linear simulations (not shown here). In a later time after the pellet injection% (seconds 4 to 6)
, the plasma relaxes to a poorer confinement state,
%which we can consider as typical from {gas-fuelled discharges in} OP1.2 campaign, 
with a flat density profile (with $n< 3.5 \times 10^{19} ~\rm{m}^{-3}$) and peaked ion and electron temperature profiles at the same time (with central values $T_i \sim 1.5 ~\rm{keV}$ and $T_e \sim 3 ~\rm{keV}$), characteristic of ECH heated discharges in OP1.2, which features a situation with ITG instability (see \citep{Bozhenkov2020}). 
The profiles used in our simulations are shown in figure 18 and are characteristic of the normal confinement phase in that discharge around t=4s. We use analytical fits of the form $X = b[1-(\frac{r}{a})^c]^d+e$, where $X$ represents $T_e$, $T_i$, or $n$. To investigate the role of collisions we run a simulation with a doubled density profile (see figure 18).

%*************************************************************************************
\begin{figure}
	\centering
	{\includegraphics[draft=false,  trim=5 0 20 15, clip, width=6.2cm]{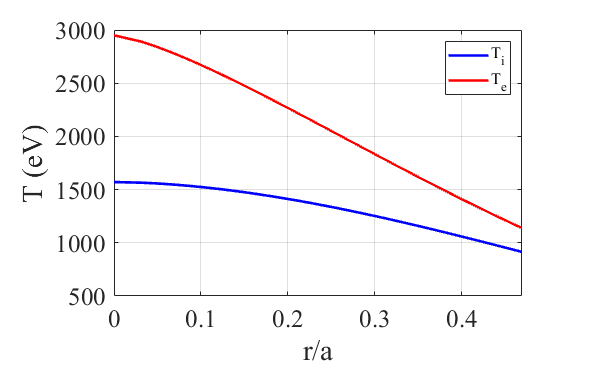}}
	{\includegraphics[draft=false,  trim=5 0 20 15, clip, width=6.2cm]{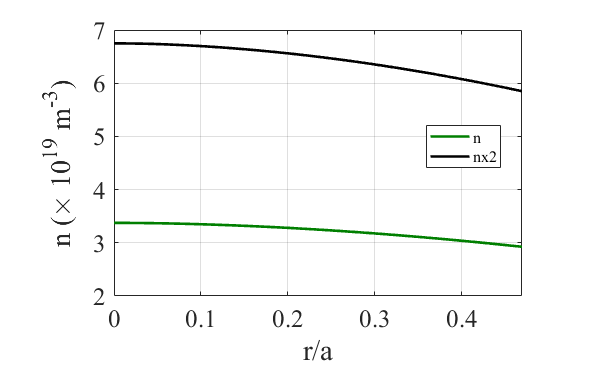}}
	\\
	\caption{Temperature and density profiles from  W7-X program 20181016\_37 \citep{Bozhenkov2020}, for times { around t=4s}. % The profiles are defined in Eq. \ref{eqNTProfilesExpW7X}. Two different profiles of density are used, the one defined in \ref{eqNTProfilesExpW7X}
		The experimental density profile and another one with doubled the density are shown.}
	\label{figNTProfilesW7XExp}
\end{figure}
%*************************************************************************************
%	We use two different profiles of density, the experimental one, %in \ref{eqNTProfilesExpW7X} 	and also a profile with twice the experimental  density, with the purpose of emphasizing the role of collisions in the simulation.% defined in that equation.
	
\begin{figure}
	\centering
	
	\includegraphics[draft=false,  trim=0 52 10 0, clip, width=6.5cm]{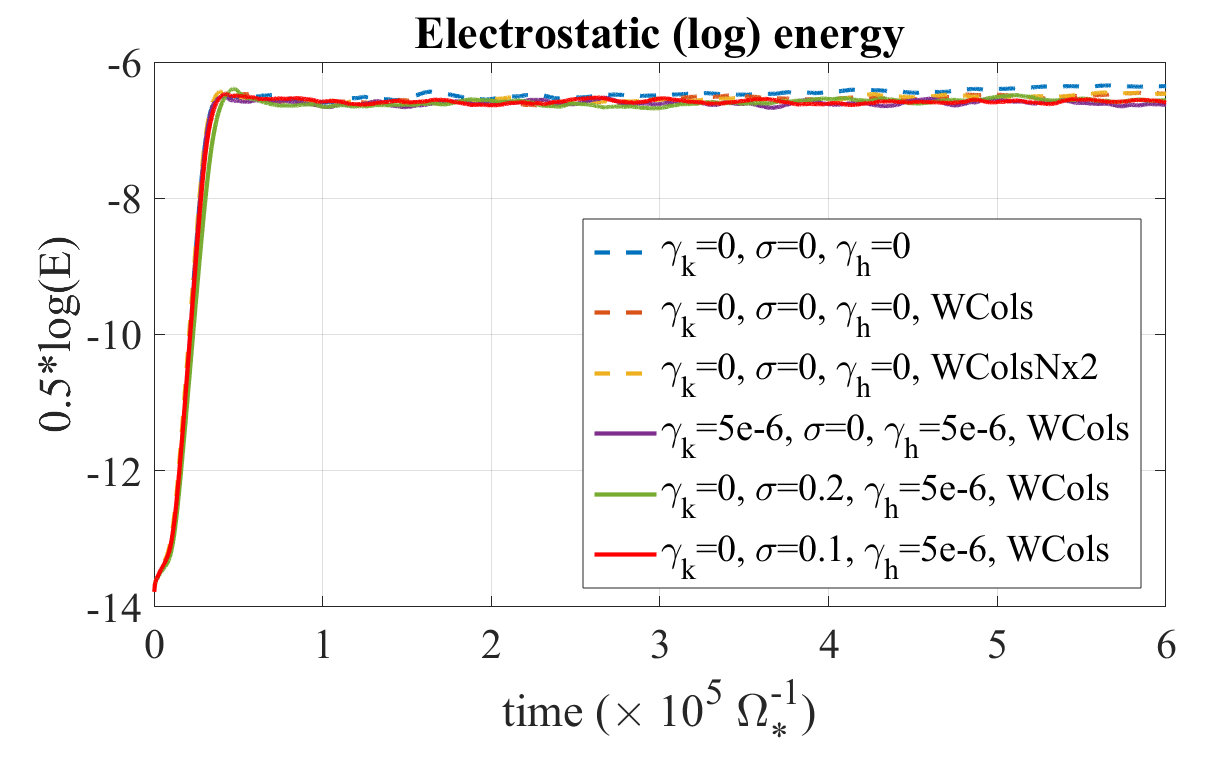} 
	\includegraphics[draft=false,  trim=0 52 10 0, clip, width=6.5cm]{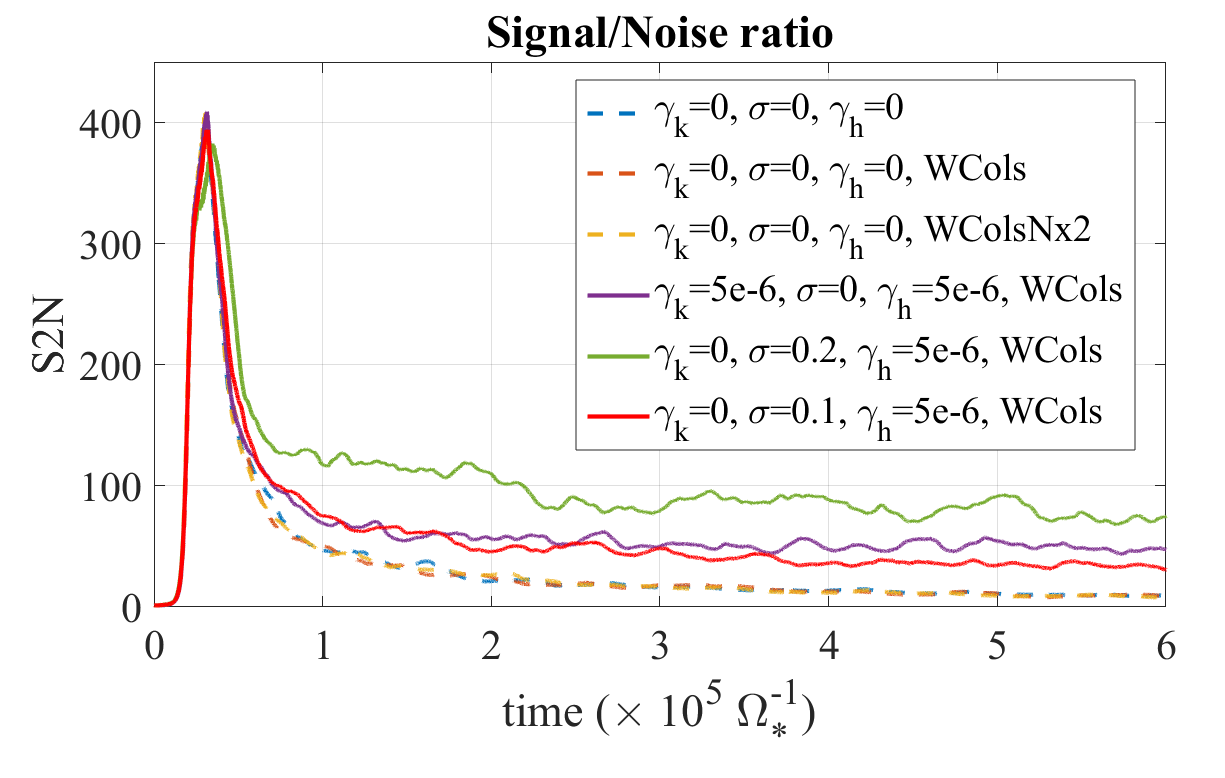}\\
	\includegraphics[draft=false,  trim=0 0 10 0, clip, width=6.5cm]{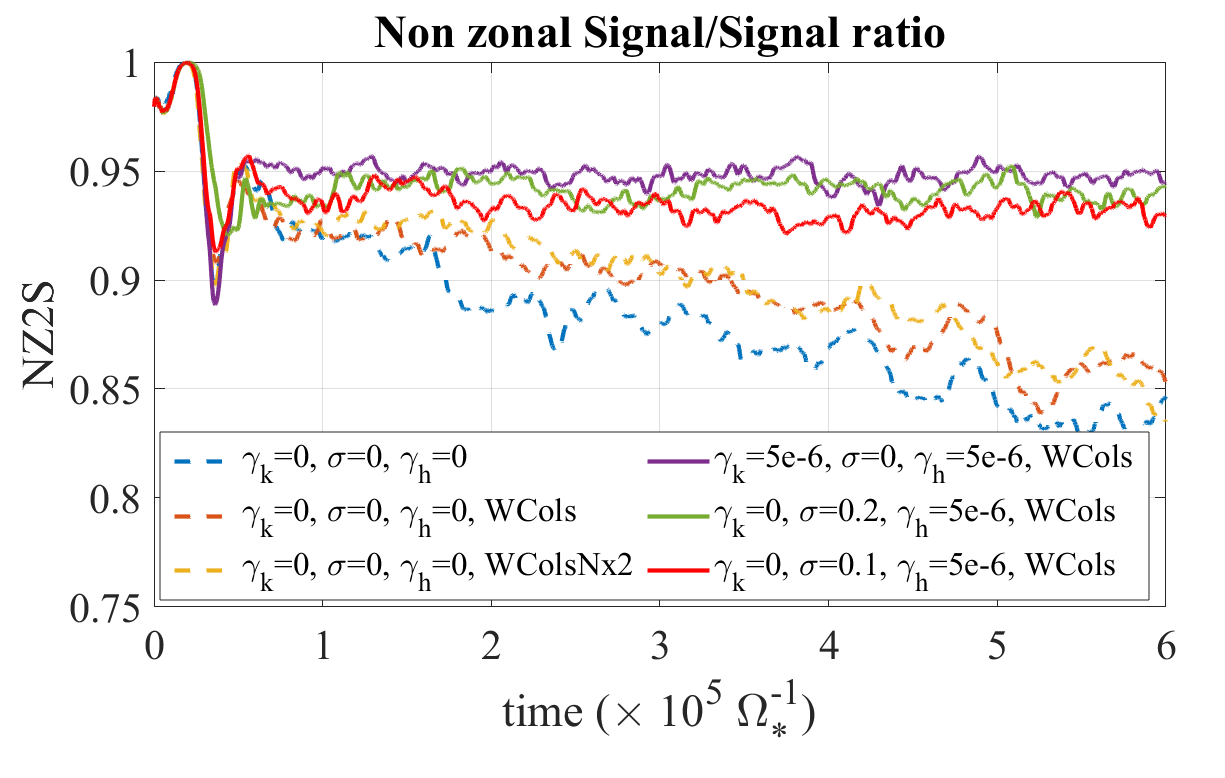}
	\includegraphics[draft=false,  trim=0 0 10 0, clip, width=6.5cm]{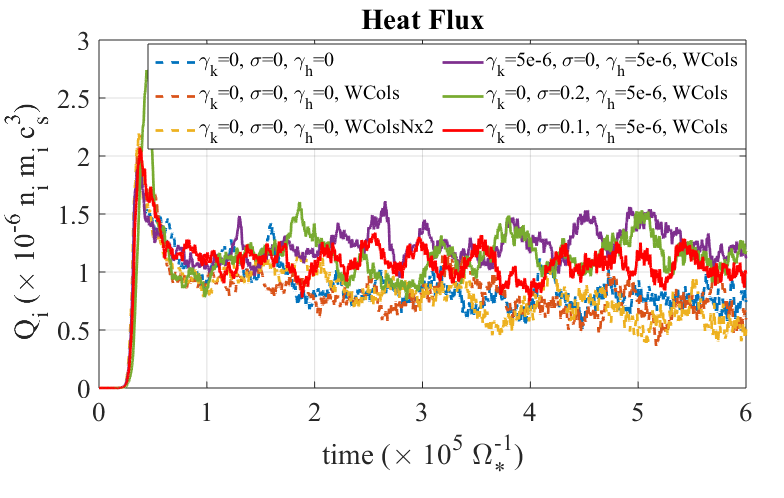}\\
	
	\caption{Electrostatic field energy (top left), S2N ratio (top right), NZ2S ratio (bottom left) and volume-integrated heat flux (bottom right) for a set of simulations using W7X- standard configuration and experimental-like density and temperature profiles.}
	\label{figResSimsW7XExp}
\end{figure}

We will apply the noise control and stabilization tools previously presented in a set of nonlinear simulations using these settings. We first run a linear simulation with these profiles without using any form of noise control. In this simulation the maximum linear growth rate is $\gamma_{max}=4 \times 10^{-4} \Omega_*$.
 Then, we have studied six different cases. First, as a reference case, we run a nonlinear simulation without accounting for collisions and without using any NCST. Next, we run another one incluiding collisions, with a pitch angle scattering collision operator \citep{Kauffmann2010,Regana2013}. Then, a simulation including collisions and using a density profile with doubled density is considered in order to artificially enhance the influence of the collisions on the stabilization of heat flux and on the numerical noise, if any.
The rest of simulations all include collisions with the experimental density profile from figure \ref{figNTProfilesW7XExp} and using several NCSTs. The first one includes a Krook operator with moderate parameter $\gamma_k=5\times 10^{-6}\Omega_*$  %, a quad tree smoothing with parameters $f_{QT}=1\Omega_*, \sigma_{QT}=0.2, N_{QT}=10$ 
and also a heating source with $\gamma_h=5\times 10^{-6}\Omega_*$. Next one is a simulation without Krook operator and using QT smoothing and heating sources, with  $f_{QT}=1\Omega_*, \sigma_{QT}=0.2, N_{QT}=10$ and $\gamma_h=5\times 10^{-6}\Omega_*$. Finally, a simulation with QT smoothing and heating, using a less agressive smoothing is considered, $f_{QT}=1\Omega_*, \sigma_{QT}=0.1, N_{QT}=10$ and $\gamma_h=5\times 10^{-6}\Omega_*$. The results are compared in figure \ref{figResSimsW7XExp}, which shows the electrostatic energy, the S2N and NZ2S ratios and the volume integrated heat flux for all these cases.
%*************************************************************************************
%*************************************************************************************
First, we can highlight the weak influence of the collisions in all the measures studied. The results including collisions, even with doubled density, are almost the same as without including them. In particular, the  S2N and NZ2S ratios are  very similar with or without collisions, but also the heat fluxes are almost the same. S2N and NZ2S significantly improve both with Krook or QT smoothing. The influence of heating alone on S2N and NZ2S is very small (not shown), as expected, but it  allows stabilizing the heat flux level. The heat flux level changes only slightly when QT smoothing is added, but the S2N quality improves significantly. Not much difference in the heat flux level is found between smoothing with $\sigma_{QT}=0.1$ or $\sigma_{QT}=0.2$. The Krook operator also improves S2N ratio with a larger influence on the heat flux, which can change depending on the value of $\gamma_k$.

After the detailed study discussed in previous sections and the results presented in this one, we can conclude that stable reliable simulations with realistic parameters can be conducted using only heating sources to sustain the profiles and a soft QT smoothing in order to improve the simulation quality and without using a Krook operator. A Krook operator can also improve the S2N ratio; however it has the drawback that can artificially affect collisions and, as demonstrated in previous sections, also affects the  zonal flow dynamics, even in the short-time. In addition, no clear saturation of the heat flux level is observed when $\gamma_k$ is increased. A heating source with a small $\gamma_h$ (a factor 100 below the maximum growth rate) parameter does not affect significantly neither the linear growth rate or the linear evolution of the zonal flow, as shown in section \ref{secInflNCSTLinZFs}. Adding a soft weight smoothing, with $\sigma \sim 0.1$, contributes to improving the S2N ratio, stabilizing the zonal component and the heat flux, while having a small effect on the short-time linear evolution of the zonal flows. A good and steady S2N ratio, stable zonal flow component and heat flux can be reached with only these two elements.

%***************************************************************************************
\section{Summary and conclusions}\label{secSumConcl}

  	In this work a comparison of global adiabatic-electron simulations carried out with the PIC codes ORB5 and EUTERPE in a tokamak configuration has been conducted. Good agreement was found in the linear properties of unstable modes between both codes. 
  	
  	Both growth rates and real frequencies were studied for different density and temperature profiles and a good agreement on the calculations of these quantities with both codes has been found.   	 
  	EUTERPE has been benchmarked against ORB5 also in a nonlinear setting by means of simulations with adiabatic electrons in the same tokamak (Cyclone Base Case) equilibrium. The simulations have been conducted with the same numerical settings in both codes and, despite of the differences between codes, a good agreement is found in the volume integrated heat flux and heat conductivity provided that sufficient statistics is used. 
  	
  	In addition, a detailed study of several tools recently implemented in EUTERPE for the stabilization of simulations and the control of numerical noise has been carried out. Three tools have been thoroughly studied: heating sources implemented to sustain the density and temperature profiles, a Krook-type source term and also a smoothing of the markers weights which are aimed at reducing the numerical noise inherent to the PIC method and stabilizing the simulations. These tools have been tested in a tokamak (CBC) equilibrium and also in the standard configuration of the W7-X stellarator. The influence of using these tools on both linear (linear growth rate of unstable modes, linear relaxation of zonal flows) and nonlinear (turbulent heat flux) properties has been studied in detail.   	
  	The influence on the linear growth rate of unstable modes has been studied in tokamak and stellarator (W7-X) configurations while for the linear zonal flow evolution the focus is put on the stellarator and the specific features (low frequency oscillation) appearing on it.
  	 
  	It is found that, as expected, the heating sources allow sustaining the profiles, with limited effect on the linear properties and signal quality. Both Krook operator and smoothing affect the linear growth rate of modes and the linear relaxation of zonal flows. An increasing reduction of the linear growth rate is observed as the frequency of the Krook source term is increased, with moderate changes for values of $\gamma_k$ one order of magnitude below the linear growth rate. The weight smoothing 
  	has also an increasing effect on the linear growth rates which can be kept small for small values of the sigma parameter.
  	While using a Krook source term produces a damping of zonal flows at all time scales which is strong even for very small values of the frequency (well below the linear growth rate), the smoothing has a very limited influence on the time evolution of ZF for times shorter than time scales typical from the turbulence (turbulence decorrelation time or eddy turn-over time) in the order of tens of microseconds. For longer times the influence is also strong, however. Then, we can conclude that quad tree weight smoothing affects the zonal flow evolution less than the Krook operator at short times, which can be argued to be the most relevant time scales for turbulence saturation.
  	
  	The influence of these tools in nonlinear simulations has also been studied. Several measures have been used to quantify the improvement of simulation quality. First, a signal to noise ratio has been used. In addition, the ratio of non-zonal component of the signal to the total signal has been used. Special attention has been paid to the volume integrated, amplitude of the zonal component of potential, which can be largely affected by the numerical noise. Finally, the influence of these tools on the volume integrated heat flux has been studied. Using heating sources allows sustaining the density and temperature profiles and then helps in obtaining a stable heat flux signal, while its effect on simulation quality, as measured by S2N or ZF2S, is very small. Using either a Krook source term or the weight smoothing allows to improve the simulation quality as measured by the S2N and ZF2S. Weight smoothing affects  slightly to the heat flux, while the influence of the Krook operator is stronger. 
  	The strong influence of Krook operator on the linear ZF damping at short times and the strong effect that it can produce on the turbulent heat flux makes the weight smoothing  a preferred means for the numerical noise control, particularly in collisional simulations in which the Krook source term can interfere and distort the effect of a more realistic collision operator. Stable simulations with improved (reduced) noise can be obtained using heating sources, with appropriate parameters, to sustain the profiles and weight smoothing for improving the simulation quality. The small damping  of zonal flows at short times allows a proper treatment of zonal flow response. The  weight smoothing presently implemented lacks of energy and momentum conservation, however. Work is in progress to develop a weight smoothing with conservation properties.
  	
  	The reliability of a radially restricted simulation has been studied in a set of simulations in which different radial domains of computation are used instead of simulating the full volume. This is important in order to reduce the computation resources for a nonlinear simulation, which can be huge if the full radial domain is considered in realistic experimental conditions.
  	
  	As an application of the tools for noise control and stabilization, a realistic case of W7-X has been studied. Radially restricted collisional nonlinear simulations using adiabatic electrons and realistic density and temperature profiles from a recent experimental program in W7-X have been carried out. 
  	The effect of NCSTs on the heat flux and quality measures has been studied in this practical case showing similar results to those observed in collisionless simulations with ideal density and temperature profiles carried out in the full radial domain. The  turbulent heat flux in this case can also be stabilized using heating sources and weight smoothing, reaching a quasi steady state with reduced numerical noise.
  	
  	 To the best of our knowledge, these are the first nonlinear simulations carried out in a stellarator with a  global particle-in-cell code, using realistic experimental plasmas parameters and reaching a turbulent saturated steady state, which have been reported.

%***************************************************************************************
\section{Acknowledgments}
%***************************************************************************************

The authors thank E. Sonnendr\"ucker for the implementation of quad tree smoothing and to the CXRS and Thomson groups  at  W7-X for providing the experimental density and temperature profiles. We thank A. Alonso, I. Calvo and J. Riemann for useful discussions and comments. 
We acknowledge the computer resources at Mare Nostrum IV and the technical support provided by the Barcelona Supercomputing Center. Part of the simulations were carried out using the Marconi supercomputer at CINECA, from the EUROfusion infraestructure.
This work has been partially funded by the Ministerio de Ciencia, Innovaci\'on y Universidades of Spain  under project PGC2018-095307-B-I00.
This work has been carried out within the framework of the EUROfusion Consortium and has received funding from the Euratom research and training programme 2014-2018 and 2019-2020 under grant agreement N$^{\rm{o}}$ 633053. The views and opinions expressed herein do not necessarily reflect those of the
European Commission.
\bibliographystyle{jpp}
% Note the spaces between the initials
\bibliography{Sanchez_EFTC19_4_JPP}

\end{document}